\documentclass[12pt,a4paper]{article}
\usepackage{latexsym,amsmath,amssymb,verbatim}
\date{}
\numberwithin{equation}{section}

\begin{document}
\title{Supersymmetric Quantum Field Theory: Indefinite Metric}
\author{Florin Constantinescu\\ Fachbereich Informatik und Mathematik \\ Johann Wolfgang Goethe-Universit\"at Frankfurt\\ Robert-Mayer-Strasse 10\\  60054
Frankfurt am Main, Germany }
\maketitle

\begin{abstract}
We study the recently introduced Krein structure (indefinite metric) of the $N=1$ supersymmetry and present the way into physical applications outside path integral methods. From the mathematical point of view some perspectives are mentioned at the end of the paper.
\end{abstract}

\section{Introduction}

It was claimed \cite{C1} that the $N=1$ superspace in four space-time dimensions hides an inherent indefinite metric which can be realized as a Krein space. It gives rise by standard methods to an invariant Hilbert space realized on supersymmetric functions (supersymmetric Hilbert space). The pair of these two spaces was called in \cite{C1} the Krein-Hilbert (or Hilbert-Krein) structure of supersymmetry. \\
In this paper we present several detailed simple proofs of this assertion together with the first steps towards the study of quantum supersymmetric fields outside path integral methods. In particular our methods open the way into less well-known subjects as for instance the canonical quantization of supersymmetries and the supersymmetric K\"allen-Lehmann representation.  \\
The paper is structured as follows: in Section 2 we describe our notations and conventions. Generally they coincide with those of \cite{WB} with only one exception (see Section 2). Our Minkowski metric is $(-1,1,1,1)$. Our study needs considerations on supersymmetric functions with commuting numerical spinor components instead of the usual supersymmetric fields used in the framework of path integral methods. In order to cope with both, functions and fields, in Section 3 we modify and enrich the methods of computations in a form which we call mixed van der Waerden calculus. Sections 4,5,6 contain preparatory material. The main point of the paper is explained in Section 7. Proofs are presented in Sections 8 and 9 and the results are evaluated in Section 10. Sections 11 and 12 contain technical as well as physical applications to the supersymmetric quantum field theory. 
Note that our Hilbert supersymmetric space is different from the super Hilbert space of \cite{DeW,DM,V}. The latter is not used in this paper. \\
We close the Introduction by some remarks which might help the reader to place the contents of this paper into an adequate framework.\\
Indefinite metric is well known in quantum physics. For example in electrodynamics it appears under the name of Gupta-Bleuler and St\"uckelberg quantization as well as quantization using ghost fields. It could be of interest in non-abelian gauge theories too. In a rigorous setting, which is unusual and therefore less known to physicists (we apologise for using a less well known approach) it appears in the form of a Krein space \cite{StW,St}. It makes no problems because an adequate Hilbert space can be easily recovered. From technical point of view the rigorous setting uses the so called axiomatic approach to quantum field theory in which quantum fields are after all suposed to be operator-valued distributions \cite{SW}. According to general rules of distribution theory, operations on fields are transfered to test functions. As a particular example, which might help understanding the present paper, we mention the well known Gupta and Bleuler subsidiary (annihilation) condition. It can be transfered to test functions and helps defining the physical positive definite Hilbert space on functions of space-time. The same procedure works in electrodynamics in the St\"uckelberg quantization or even quantization using ghost fields (see for instance \cite{Sch} Chapter 1 or the review paper \cite{G}). 
In this paper we first work out in detail the intrinsic indefinite metric of the $ N=1 $ superspace claimed in \cite{C1} in the massive case and start applications of this structure (physical and mathematical) keeping in mind the analogy with the above discussion of indefiniteness in quantum electrodynamics. Our considerations will not be limited to free fields (see the last section of the paper). Test functions and distributions can be extended to super test functions and super distributions but the Hilbert space keeps its meaning exactly, being now realized on super functions (Hilbert superspace). As already mentioned above it is different from the so-called super Hilbert space used sometimes in supersymmetry (which is neither a Krein nor a Hilbert space) in which the indefiniteness is more stringent than in the Krein space itself because it contains vectors of imaginary lengts. Following the paper the reader will find out that our methods are appropriate for the relativistic case but not for the supersymmetric quantum mechanics. 

\section{Notations and Conventions}

The signature of Minkowski space is $(-1,1,1,1)$. Associated to x in Minkowski space there are two-component Grassmann variables $\theta, \bar \theta $. Basically we use common notations and conventions following \cite{WB} with only one exception concerning the sign of $\sigma^0 $ which is one in this paper instead of minus one in \cite{WB}. They fully coincide with the notations in \cite{S}. 
We make difference between supersymmetric functions, supersymmetric fields and quantum supersymmetric fields. The supersymmetric functions simulate (up to regularity properties) both wave functions as well as test functions. Supersymmetric fields, which are common in physical textbooks, are milestones of path integrals. Because our work lies outside path integrals, they will not be really used in this paper (except for some side remarks). Supersymmetric (test) functions of $z=(x,\theta, \bar \theta );\theta =(\theta_{\alpha}),\alpha=1,2 ;\bar \theta =(\bar \theta_{\dot \alpha }),\dot \alpha =\dot 1,\dot 2 $ are written as
\begin{gather}\nonumber 
X(z)=X(x,\theta ,\bar \theta )= \\ \nonumber
=f(x)+\theta \varphi (x) +\bar \theta \bar \chi (x) +\theta ^2m(x)+\bar \theta^2n(x)+ \\ 
+\theta \sigma^l\bar \theta v_l(x)+\theta^2\bar \theta \bar \lambda(x)+\bar \theta^2\theta \psi (x)+ \theta^2 \bar \theta^2d(x)
\end{gather}
where for definitness we choose the coefficients of $\theta ,\bar \theta $ to be regular functions of $x$ decreasing to zero at infinity (for instance in the Schwartz space $S$) but eventually we will allow some singularities (distributions). The coefficients of odd powers of Grassmann variables are numerical spinors i.e. spinors with numerical components. Bar means complex conjugation for numbers and functions e.g. $\bar \varphi_{\dot \alpha }=\overline{\varphi_\alpha }=(\varphi_{\alpha })^* $ as well as conjugation for the Grassmann variables e.g. $\bar \theta_{\dot \alpha}=(\theta_{\alpha})^* $. The variables $\theta ,\bar \theta $ are looked at as independent. For $v_l(x)$ we can write equivalently
\begin{gather}
\theta \sigma^l\bar \theta v_l(x)=\theta ^{\alpha }\bar \theta ^{\dot \beta }v_{\alpha \dot \beta }(x) 
\end{gather}
where 
\begin{gather}
v_{\alpha \dot \beta }(x)=\sigma _{\alpha \dot \beta }^l v_l(x) 
\end{gather}
and other way round
\begin{gather}
v^l(x)=-\frac{1}{2}\bar \sigma^{l\dot \beta
\alpha }v_{\alpha \dot
  \beta }(x) 
\end{gather}
This shows that the components of the Grassmann variables in (2.1) are arbitrary functions of $x$ (in $S$) as they should be. As stated before $x$, or more precisely its components $x^l ,l=0,1,2,3 $, are numbers (base) but eventually we will be forced to admit that they acquire even elements of the Grassmann algebra. In this case we tend to perform the Taylor expansion retaining for $x$ only the base. \\
We consider the complex linear space of supersymmetric functions of type (2.1). Eventually we will consider supersymmetric functions and distributions of several space-time and Grassmann variables too. We do not undertake any effort in order to mathematically define functions of several variables (especially Grassmann); instead we invoke at this point the handwaveing arguments in physics. \\
Usually expressions of form (2.1) used in physics where the coefficients of the odd powers of the Grassmann variables are spinors with anticommuting components (which anticommute with $\theta, \bar \theta $ too) are called (classical) fields and appear in the process of supersymmetric path integral quantization. We will encounter supersymmetric fields in this paper only marginally. \\
In what follows we use a mixed van der Waerden calculus which takes into account as usual the anticommutativity of the components of $\theta, \bar \theta $ as well as the commutativity of the numerical spinor components among them and with Grassmann's $\theta, \bar \theta $ too. For the convenience of the reader, in the next section, we give a full account of computational rules for the mixed van der Waerden calculus.

\section{Mixed van der Waerden Calculus}

The standard van der Waerden calculus \cite{WB,S} turns spinor matrix algebra into a spinor tensor calculus common for vectors and tensors. It is used in supersymmetry together with an overall convention of anticommutativity of the Grassmann variables and spinor components.  The main tools are the antisymmetric "metric tensors" $(\epsilon^{\alpha \beta }),(\epsilon_{\alpha \beta }),\epsilon_{21}=\epsilon^{12}=1,\epsilon_{12}=\epsilon^{21}=-1  $ and the $\bar \sigma $ matrix
\begin{gather}
\bar \sigma^{l\dot \alpha \alpha }=\epsilon^{\dot \alpha \dot \beta }\epsilon^{\alpha \beta }\sigma_{\beta \dot \beta }^l \quad l=0,1,2,3
\end{gather}
The matrices $\sigma_{\alpha \dot \alpha }^l,l=1,2,3 $ are the usual Pauli matrices:
\[ \sigma^1 =\begin{pmatrix}
0&1 \\
1&0 
\end{pmatrix}, 
\sigma^2 =\begin{pmatrix}
0&-i \\
i&0
\end{pmatrix},
\sigma^3 =\begin{pmatrix}
1&0 \\
0&-1
\end{pmatrix} \] 
and $\sigma^0 =1$. Here $\epsilon_{\dot \alpha \dot \beta }=\overline {\epsilon_{\alpha \beta }},\epsilon^{\dot \alpha \dot \beta }=\overline {\epsilon^{\alpha \beta }}$. For the Kronecker symbol $\overline{\delta_\alpha ^\beta }=\delta_{\dot \alpha }^{\dot \beta }$. We have $\epsilon_{\alpha \beta }\epsilon^{\beta \gamma }=\delta_{\alpha }^{\gamma },\epsilon_{\dot \alpha \dot \beta }\epsilon^{\dot \beta \dot \gamma }=\delta_{\dot \alpha }^{\dot \gamma } $. Note the standard index positions for $\sigma $ and $\bar \sigma $ which make contact to the matrix interpretation: up for $\bar \sigma $, down for $\sigma $. Eventually when reading up some Hilbert space properties from the formal van der Waerden calculus to follow this caution will be important. In matrix form we have $\bar \sigma =(\sigma^0 ,-\sigma^1 ,-\sigma^2 ,-\sigma^3 )$. \\
Spinors $ \psi $ with upper and lower indices are related through the $\epsilon $-tensor:
\begin{gather}
\psi^\alpha =\epsilon^{\alpha \beta }\psi_\beta , \psi_\alpha =\epsilon_{\alpha \beta }\psi^\beta
\end{gather} 
On the way we accept in some places nonstandard index positions for $\sigma, \bar \sigma $ obtained with the help of $\epsilon ,\bar \epsilon $ as in \cite{S} but return to standard positions in the end of the computation, especially when we connect to the matrix interpretation. Nonstandard index positions in $\sigma , \bar \sigma $ may easily produce confusions and therefore they must be used with care.\\  
We have
\begin{gather}
\bar \sigma_{\dot \alpha \beta }^l =\sigma_{\beta \dot \alpha }^l =\overline{\sigma_{\alpha \dot\beta }^l} =(\sigma_{\alpha \dot \beta}^l )^* ,\quad \overline{\bar \sigma_{\dot \alpha \beta }^l }=\sigma_{\alpha \dot\beta }^l
\end{gather}
Care must be paid also to the bar on $\sigma $ which has double meaning but is clear from the context. For example in $ \overline{\bar \sigma_{\dot \alpha \beta }^l }=\sigma_{\alpha \dot\beta }^l $ the upper bar means complex conjugation. These relations are compatible with (3.1).
We also write for instance
\[\bar \sigma_{\dot \alpha }^{l\beta }=\sigma_{\dot \alpha}^{l\beta }=\overline{\sigma_\alpha ^{l\dot \beta }} \]
i.e. knowing that always for $\sigma $ the first index is understood to be undotted whereas the second one is dotted and vice-versa for $\bar \sigma $ though typographically in our notations this is not visible for $\sigma_{\dot \alpha }^{l\beta } ,\bar \sigma_{\dot \alpha }^{l\beta },\sigma_{\alpha }^{l\dot \beta } $. Note that the complex conjugation bar changes indices (as for instance in (3.3)). \\
We have for $ l,m=0,1,2,3 $
\begin{gather}
tr(\sigma^l \bar \sigma^m )=\sigma_{\alpha \dot \beta }^l \bar \sigma^{m\dot \beta \alpha} =-2\eta^{lm} \\ 
\sigma_{\alpha \dot \beta }^l \bar \sigma_l ^{\dot \rho \sigma } =-2\delta_{\alpha }^{\sigma }\delta_{\dot \beta }^{\dot \rho } \\ 
(\sigma^l \bar \sigma^m +\sigma^m \bar \sigma^l)_{\alpha }^{\beta }=-2\eta^{lm}\delta_\alpha ^\beta \\ 
(\bar \sigma^l \sigma^m +\bar \sigma^m \sigma^l)_{\dot \beta }^{\dot \alpha }=-2\eta^{lm}\delta_{\dot \beta }^{\dot \alpha }
\end{gather}
where $ \eta $ is the Minkowski metric tensor $(-1,1,1,1)$.
In the anticommuting case i.e. in the standard van der Waerden calculus we have
\begin{gather}\nonumber
\psi \chi =\psi^\alpha \chi_\alpha =-\psi_\alpha \chi^\alpha =\chi^\alpha \psi_\alpha =\chi \psi \\ \nonumber
\bar \psi \bar \chi =\bar \psi^{\dot \alpha }\bar \chi_{\dot \alpha }=-\bar \psi_{\dot \alpha }\bar \chi^{\dot \alpha }=\bar \chi^\alpha \bar \psi_\alpha =\bar \chi \bar \psi
\end{gather} 
The conjugation $(\chi_\alpha )^*=\bar \chi_{\dot \alpha }, (\chi^\alpha )^*=\bar \chi^{\dot \alpha } $ reverses the order of spinor components: 
\begin{gather}\nonumber
(\chi \psi )^*=\overline {\chi \psi }=\overline {\chi^\alpha \psi_\alpha }=\bar \psi_{\dot \alpha }\bar \chi^{\dot \alpha }=\bar \psi \bar \chi =\bar \chi \bar \psi =\overline {\psi \chi }=(\psi \chi )^*
\end{gather}
We also have the usual list of rules in the (anticommuting) spinor algebra given in Appendix A and Appendix B of \cite{WB}:
\begin{gather}\nonumber
(\theta \phi )(\theta \psi )=-\frac{1}{2}(\phi \psi )\theta^2 \\ \nonumber
(\bar \theta \bar \phi )(\bar \theta \bar \psi )=-\frac{1}{2}(\bar \phi \bar \psi )\bar \theta^2 \\ \nonumber
\chi \sigma^l \bar \psi =-\bar \psi \bar \sigma^l \chi \\ \nonumber
\overline {\chi \sigma^l \bar \psi }=\psi \sigma^l \bar \chi
\end{gather} 
We use notations like $\varphi \sigma ^l , \sigma^l\bar \varphi , \bar \varphi \bar \sigma^l $ etc. meaning the spinors $(\varphi \sigma^l )_{\dot \beta }=\varphi^{\alpha } \sigma_{\alpha \dot \beta }^l , (\sigma^l \bar \varphi )_\alpha =\sigma_{\alpha \dot \beta }^l \bar \varphi^{\dot \beta }, (\bar \varphi \bar \sigma^l )^\alpha =\bar \varphi_{\dot \beta }\bar \sigma^{l \dot \beta\alpha }, (\bar \sigma^l \varphi )^{\dot \beta }=\bar \sigma^{l\dot \beta \alpha }\varphi_\alpha $ i.e. respecting the standard index positions in $\sigma $. For example
\[(\bar \chi \bar \sigma^l )\psi =(\bar \chi \bar \sigma^l )^\alpha \psi_\alpha =\bar \chi_{\dot \alpha } \bar \sigma^{l\dot \alpha \alpha } \psi_\alpha =\bar \chi (\bar \sigma^l \psi )=\bar \chi \bar \sigma^l \psi     \]
Concerning the differential and integral calculus in the Grassmann variables we follow usual conventions too (see for instance \cite{S}). The Grassmann derivatives are
\[ \partial_{\alpha}=\frac {\partial }{\partial {\theta^{\alpha }}},  \partial^{\alpha}=\frac {\partial }{\partial {\theta_{\alpha }}}, \bar \partial_{\dot \alpha}=\frac {\partial }{\partial {\bar \theta^{\dot \alpha }}},  \bar \partial^{\dot \alpha}=\frac {\partial }{\partial {\bar \theta_{\dot \alpha }}}          \]
defined through
\begin{gather}\nonumber
\partial_{\alpha}\theta^{\beta }=\delta_{\alpha }^{\beta }, \partial^{\alpha}\theta_{\beta }=\delta_{\beta }^{\alpha },\bar \partial_{\dot \alpha}\bar \theta^{\dot \beta }=\delta_{\dot \alpha }^{\dot \beta }, \bar \partial^{\dot \alpha}\bar \theta_{\dot\beta }=\delta_{\dot \beta }^{\dot \alpha } \\ \nonumber
\bar \partial_{\dot \alpha }\theta^{\beta }=\partial_{\alpha }\bar \theta^{\dot \beta }=0,  \{ \partial_{\alpha},\partial_{\beta }\}  = \{\bar \partial_{\dot \alpha },\bar \partial_{\dot \beta }\}  =0 
\end{gather}          
Derivatives of products of Grassmann variables are defined using the product rule where the derivatives anticommute with the variables. In particular
\[\partial_{\alpha }\theta^2 =2\theta_{\alpha }, \bar \partial_{\dot \alpha }\bar \theta^2 =-2\bar \theta_{\dot \alpha }, \partial^{\alpha }\theta^2 =-2\theta^{\alpha }, \bar \partial^{\dot \alpha }\bar \theta^2 =2\bar \theta^{\dot \alpha }\]
where $ \theta^2 =\theta \theta ,\bar \theta^2 =\bar \theta \bar \theta $. \\
For derivatives we have
\begin{gather}\nonumber
\epsilon^{\alpha \beta }\partial_{\beta} =-\partial^\alpha,\epsilon_{\alpha \beta }\partial^{\beta} =-\partial_\alpha,\epsilon^{\dot \alpha \dot \beta }\bar \partial_{\dot \beta} =-\bar \partial^{\dot \alpha }, 
\epsilon_{\dot \alpha \dot \beta }\bar \partial^{\dot\beta} =-\bar \partial_{\dot \alpha }
\end{gather}
The Grassmann derivative is an operator. In order to connect to standards in physics we will introduce the conjugation of operators but not use it in this paper. The conjugation of the Grassmann derivative is defined using the left derivative to be \cite{S} 
\begin{gather}\nonumber
(\partial_\alpha )^*=\overline{\partial_\alpha }=\overleftarrow{\bar \partial }_{\dot \alpha }
\end{gather}
It can be proved that
\begin{gather}
(\partial_\alpha )^*=\overline{\partial_\alpha }=\mp \bar \partial_{\dot \alpha } 
\end{gather}
where the signs appear if we apply the derivative to an even or to an odd function of the Grassmann variables.
This is at the same level of formality as the definition of conjugation of the usual derivative: $ (i\frac {\partial        }{\partial x^l })^* =-i\frac {\partial }{\partial x^l } $ which has to be contrasted with the definition of the $L^2 $ (Hilbert) -adjoint operator $ (i\frac {\partial }{\partial x^l })^{\dagger } =i\frac {\partial }{\partial x^l } $ . Later on we will discuss supersymmetric adjoint operators in a Hilbert space to be still defined. The above conjugation will play no role in this paper although it could be used to give quick alternative proofs of some relations to follow in Section 5. We define finally $\partial^2 =\partial^{\alpha }\partial_\alpha ,\bar {\partial^2 }=\bar \partial_{\dot \alpha }\bar \partial^{\dot \alpha }$.  \\
Some particular aspects of Berezin integration in our context i.e. in the presence of complex and Grassmann conjugation, will be discussed later on in the paper.\\
As long as we work in physics with supersymmetric fields the standard (anticommutative) van der Waerden calculus described above is sufficient and very useful. But if we want to apply it to supersymmetric functions of the form (2.1) with numerical spinor coefficients the rules have to be modified and enriched. We describe now these enrichments. \\
Suppose that the spinor components are assumed to commute between themselves and with the Grassmann variables $\theta, \bar \theta $. Up to this modification we retain all conventions from above. Obviously the anticommutativity property of the Grassmann variables and the van der Waerden rules for them remain unchanged. We start by giving the corresponding counterpart of the rules above for the commutative spinor algebra. They are:
\begin{gather}\nonumber
\psi \chi =\psi^\alpha \chi_\alpha =-\psi_\alpha \chi^\alpha =-\chi^\alpha \psi_\alpha =-\chi \psi \\ \nonumber
\bar \psi \bar \chi =\bar \psi^{\dot \alpha }\bar \chi_{\dot \alpha }=-\bar \psi_{\dot \alpha }\bar \chi^{\dot \alpha }=-\bar \chi^{\dot \alpha }\bar \psi_{\dot \alpha }=-\bar \chi \bar \psi \\ \nonumber
\overline {\chi \psi }=\overline {\chi^\alpha \psi_\alpha }=\bar \chi^{\dot \alpha }\bar \psi_{\dot \alpha }=\bar \psi_{\dot \alpha }\bar \chi^{\dot \alpha }=\bar \psi \bar \chi =-\bar \chi \bar \psi =-\overline {\psi \chi } \\ \nonumber
(\theta \phi )(\theta \psi )=\frac{1}{2}(\phi \psi )\theta^2 \\ \nonumber
(\bar \theta \bar \phi )(\bar \theta \bar \psi )=\frac{1}{2}(\bar \phi \bar \psi )\bar \theta^2 \\ \nonumber
\chi \sigma^l \bar \psi =\bar \psi \bar \sigma^l \chi \\ \nonumber
\overline {\chi \sigma^l \bar \psi }=\psi \sigma^l \bar \chi  
\end{gather}
The $ \theta $-spinor algebra relations remain unchanged, e.g.
\begin{gather}\nonumber
\theta^\alpha \theta^\beta =-\frac{1}{2}\epsilon^{\alpha \beta }\theta^2 , \quad \theta_\alpha \theta_\beta =\frac{1}{2}\epsilon_{\alpha \beta }\theta^2 \\ \nonumber
\bar \theta^{\dot \alpha }\bar \theta^{\dot \beta }=\frac{1}{2}\epsilon^{\dot \alpha \dot \beta }\bar \theta^2, \quad \bar \theta_{\dot \alpha }\bar \theta_{\dot \beta }=-\frac{1}{2}\epsilon_{\dot \alpha \dot \beta }\bar \theta^2 \\ \nonumber
(\theta \sigma^l \bar \theta )(\theta \sigma^m \bar \theta )=-\frac{1}{2}\theta^2 \bar \theta^2 \eta^{lm } 
\end{gather}
as well as for Grassmann conjugation
\[ \overline{\theta_{1\alpha }\theta_{2\beta }\ldots  \theta_{n\gamma }}= \bar \theta_{n\dot \gamma }\ldots \bar \theta_{2\dot \beta }\bar \theta_{1\dot \alpha } \]
We see that there are sign discrepancies to the anticommutivity convention. In particular the conjugate function $\bar X(z)$:
\begin{gather} \nonumber
X^* =\bar X=\bar X(x,\theta ,\bar \theta )=\\ \nonumber
=\bar f(x)-\theta \chi (x) -\bar \theta \bar \varphi (x) +\theta ^2 \bar n(x)+\bar \theta^2 \bar m(x)+ \\ 
+\theta \sigma^l \bar \theta \bar v_l(x)-\theta^2 \bar \theta \bar \psi(x)-\bar \theta^2\theta \lambda (x)+ \theta^2 \bar \theta^2 \bar d(x)
\end{gather}
is different from the conjugate field as it appears in physics textbooks:
\begin{gather} \nonumber
X^* =\bar X=\bar X(x,\theta ,\bar \theta )=\\ \nonumber
=\bar f(x)+\theta \chi (x) +\bar \theta \bar \varphi (x) +\theta ^2 \bar n(x)+\bar \theta^2 \bar m(x)+ \\ 
+\theta \sigma^l \bar \theta \bar v_l(x)+\theta^2 \bar \theta \bar \psi(x)+\bar \theta^2\theta \lambda (x)+ \theta^2 \bar \theta^2 \bar d(x)
\end{gather}
In both cases $\bar {\bar X }=X $. The term $\theta \sigma^l \bar \theta \bar v_l =\overline{\theta \sigma^l \bar \theta v_l }$ can also be written as follows
\[\theta \sigma^l \bar \theta \bar v_l =\theta^\alpha \bar \theta^{\dot \beta }\bar v_{\alpha \dot \beta }=\overline{\theta \sigma^l \bar \theta v_l }=\overline{\theta^\beta \bar \theta^{\dot \alpha }v_{\beta \dot \alpha }}=\theta^\alpha \bar \theta^{\dot \beta }\overline{ v_{\beta \dot \alpha }}        \]
which implies
\begin{gather}
\bar v_{\alpha \dot \beta }=\overline{v_{\beta \dot \alpha }}=(v_{\beta \dot \alpha })^* 
\end{gather}
as well as $\bar v_{\dot \beta }^\alpha =\overline{v_\beta ^{\dot \alpha }} $ etc. where indices are moved (up and down) with the help of $\epsilon ,\bar \epsilon $. 
In $v$ and in $\bar v $ too undotted indices are on the first and dotted indices on the second place (the situation is different from $\sigma ,\bar \sigma $ ).\\
Note that the transition from $X$ to $\bar X $ in the commuting case requires the following replacements: $f, \varphi ,\bar \chi , m,n,v,\bar \lambda ,\psi ,d $ go to $\bar f, -\chi ,-\bar \varphi ,\bar n,\bar m,\bar v,-\bar \psi ,-\lambda ,\bar d $.

\section{Some useful relations}

Up to now supersymmetry i.e. the symmetry under the Poincare supergroup (or superalgebra) played no role. In this section we prepare some tools in superspace connected to supersymmetry.
Let us consider the supersymmetric covariant (and invariant) \cite{WB,S} derivatives $D,\bar D$ with spinorial
components $D_{\alpha },D^{\alpha },\bar D_{\dot \alpha },\bar D^{\dot \alpha }$ defined as
\begin{gather}
D_{\alpha }=\partial_{\alpha } +i\sigma_{\alpha \dot \beta }^l\bar \theta^{\dot \beta }\partial _l \\ 
D^{\alpha
}=\epsilon ^{\alpha \gamma }D_{\gamma }=-\partial ^{\alpha }+i\sigma^{l\alpha }_{\dot \beta }\bar \theta^{\dot
\beta }\partial _l  \\ 
\bar D_{\dot \alpha }=-\bar {\partial }_{\dot \alpha } -i\theta ^{\beta }\sigma _{\beta
\dot \alpha }^l\partial _l \\ 
\bar D^{\dot \alpha }=\epsilon ^{\dot \alpha \dot \gamma }\bar D_{\dot
  \gamma }=\bar {\partial }^{\dot \alpha }-i\theta ^{\beta }\sigma
_{\beta }^{l\dot \alpha }\partial _l
\end{gather}
We accept here like in Section 3 nonstandard index positions for $\sigma $ writing
\[\epsilon^{\alpha \beta }\sigma^l _{\beta \dot \alpha }=\sigma^{l\alpha
}_{\dot \alpha }  \] 
Third powers monomials of $D$ -operators as well as of $\bar D$ -operators vanish. For other properties of the $D,\bar D$ -operators see \cite{S}. 
Note that $\bar D_{\dot \alpha }, \bar D^{\dot \alpha }$ were explicitly defined by (4.3),(4.4) (not by operator conjugation). This remark applies to all bar-operators to be introduced below (see also the comment before (3.8)).
Besides covariant derivatives we need supersymmetric generators \cite{S} (the definition in \cite{WB} differs by an unit imaginary factor) $Q_\alpha ,\bar Q_{\dot \alpha }$ defined as
\begin{gather}
iQ_{\alpha }=\partial_{\alpha } -i\sigma_{\alpha \dot \beta }^l\bar \theta^{\dot \beta }\partial _l \\ i\bar
Q_{\dot \alpha }=-\bar {\partial }_{\dot \alpha } +i\theta ^{\beta }\sigma _{\beta \dot \alpha }^l\partial _l
\end{gather}
satisfying
the anticommuting relations of the Poincare superalgebra \cite{WB,S}
\begin{gather}
\{Q_\alpha ,\bar Q_{\dot \beta }\}=2\sigma_{\alpha \dot \beta }^l P_l    \\
\{Q_\alpha ,Q_\beta \}=0,\{\bar Q_{\dot \alpha } ,\bar Q_{\dot \beta }\}=0
\end{gather}
where $P_l $ is the generator of space-time translations realized on functions as $P_l =-i\partial_l $.
The components of $Q,\bar Q $ commute with those of $D, \bar D$. Formally  one obtains $iQ,i\bar Q $ from $D,\bar D$ by changing $\partial_l $ to $-\partial_l $ or $ \sigma $ to $-\sigma $.
Note that $D_{\alpha }$ does not contain
the variables $\theta $ and $\bar D ^{\dot \alpha } $ does not contain the variables $\bar \theta $ such that we can easily
write at the operator level:
\begin{gather}
D^2=D^{\alpha }D_{\alpha }=-(\partial ^{\alpha }\partial _{\alpha }-2i\partial _{\alpha \dot \alpha }\bar \theta
^{\dot \alpha } \partial ^{\alpha }  +\bar \theta^2 \square ) \\ \bar D^2=\bar D_{\dot \alpha }\bar D^{\dot \alpha
}=-(\bar {\partial } _{\dot \alpha }\bar {\partial }^{\dot \alpha }+2i \theta ^{\alpha }\partial _{\alpha \dot
\alpha }\bar \partial ^{ \dot \alpha } +\theta^2\ \square )
\end{gather}
where
\[\square =\eta ^{lm}\partial _l \partial _m \]
is the d'Alembertian, $\eta $ is the Minkowski metric tensor (in our case (-1,1,1,1)) and
\[ \partial _{\alpha \dot \alpha }=\sigma _{\alpha \dot \alpha }^l \partial _l \]
We make use of operators defined as 

\begin{gather}
c=\bar D^2D^2, a=D^2\bar D^2, T=D^{\alpha }\bar D^2 D_{\alpha }=\bar D_{\dot \alpha }D^2 \bar D^{\dot \alpha
}=-8\square +\frac {1}{2}(c+a)
\end{gather}
which are used to define formal projections \cite{WB,S}
\begin{gather}
P_c=\frac {1}{16\square }c,P_a=\frac {1}{16\square }a, P_T=-\frac {1}{8\square }T
\end{gather}
on chiral, antichiral and transversal supersymmetric functions (to be rigorously defined below). These operators are, for the time being,
formal because they contain the d'Alembertian in the denominator. Problems with the d'Alembertian in (4.12) in the
denominator will be explained later in this paper but, if we wish, for the time being we may make sense of $ P_i ,i=c,a,T $ when applied to functions which in momentum space vanish in a small neighborhood of the zero momentum. When applied to such functions they are well defined in momentum space and as such in the coordinate space too. There is an alternative way to look at the inverse d'Alembertians (see the assumption after (5.40) in Section 5). Note that $ \bar c ,\bar a , \bar T ,\bar P_c , \bar P_a ,\bar P_T $ were not defined because we do not need to define them. \\
Chiral, antichiral and transversal functions are linear subspaces of general
supersymmetric functions which are defined by the conditions \cite{WB,S}
\[ \bar D^{\dot \alpha }X=0, \dot \alpha =1,2,;  D^{ \alpha }X=0, \alpha =1,2; D^2 X=\bar D^2 X=0 \]
respectively. It can be proved that these relations are formally equivalent to the relations
\[P_cX=X, \quad P_aX=X, \quad P_TX=X \] 
respectively. The index $c$ stays for chiral, $a$ for antichiral and $T$ for transversal. We have formally
\[ P_i^2=P_i, P_iP_j=0,\quad i\ne j;i,j=c,a,T \]
and $P_c +P_a +P_T =1 $. Accordingly, each supersymmetric function can be formally decomposed into a sum of a
chiral, antichiral and transversal contribution (from a rigorous point of view this statement may be wrong and has
to be reconsidered because of the problems with the d'Alembertian in the denominator; fortunately we will not run
into such difficulties as this will be made clear later in the paper). It turns out that central for our study will be the operator
\begin{gather}
J=P_c +P_a -P_T
\end{gather}
which is no longer a projection but $J^2 =1 $. $J$ is also non-local because it involves the inverse d'Alembertian.
For several purposes we also need \cite{WB}
\begin{gather}
P_+ =\frac{1}{4\sqrt {\square }}D^2 \\
P_- =\frac{1}{4\sqrt {\square }}\bar D^2
\end{gather}
They are not projection operators as the notation might suggest. \\
Let us now specify the coefficient functions
in (2.1) for the chiral $X_c $, antichiral $X_a $ and transversal $X_T $ supersymmetric functions \cite{WB,S} (they also can be read up from the formulas of the next section). \\ 
For the chiral case $X_c$ we
have:
\begin{gather} 
\bar \chi =\psi =n=0 , v_l=\partial_l (if)=i\partial_l f, \bar \lambda =\frac{i}{2}\bar \sigma^l \partial _l \varphi ,
  d=\frac{1}{4}\square f
\end{gather}
Here $ f,\varphi $ and $ m $ are arbitrary functions. \\ 
For the antichiral $X_a$ case:
\begin{gather} \nonumber
\varphi =\bar \lambda = m=0,  v_l=\partial_l (-if)=-i\partial_l f , \\ \psi =\frac{i}{2}\sigma^l
\partial_l \bar \chi ,  d=\frac{1}{4}\square f
\end{gather}
Here $f,\bar \chi $ and $ n $ are arbitrary functions. \\ 
For the transversal case $X_T$ \cite{S}:
\begin{gather} \nonumber
m=n=0, \partial_l v^l =0, \\ \bar \lambda =-\frac{i}{2}\bar \sigma^l
\partial_l \varphi ,\psi  =-\frac{i}{2}\sigma^l \partial_l \bar
\chi  , d=-\frac{1}{4}\square f
\end{gather}
Here $f,\varphi ,\bar \chi $ are arbitrary and $v$ satisfies $\partial_lv^l=0 $. \\
It is important to stress that in the above relations, for instance in (4.16), we used $\bar \lambda =\frac{i}{2}\bar \sigma^l \partial _l \varphi $ for 
$\bar \lambda^{\dot \alpha } =\frac{i}{2}(\bar \sigma^l \partial_l \varphi )^{\dot \alpha }=\frac{i}{2}\bar \sigma^{l \dot \alpha \beta}\partial_l \varphi_{\beta } $
and $\psi =\frac{i}{2}\sigma^l \partial_l \bar \chi $ for
$\psi_{\alpha }=\frac{i}{2}\sigma_{\alpha \dot \beta }^l \partial_l \bar \chi^{\dot \beta }  $
i.e. we read up starting with standard index positions in $\sigma ,\bar \sigma $. In the same vain these relations are equivalent to $ \bar \lambda =-\frac{i}{2}\partial_l \varphi \sigma^l $ or $ \lambda =\frac{i}{2}\sigma^l \partial_l \bar \varphi $ meaning 
$\bar \lambda_{\dot \alpha } =-\frac{i}{2}\partial_l \varphi^{\beta } \sigma_{\beta \dot \alpha }^l $ and $\lambda_{\alpha }  =\frac{i}{2}\sigma_{\alpha \dot \beta }^l \partial_l \bar \varphi^{\dot \beta } $ respectively etc. 
\\
Note that if $\square f ,\square \varphi_\alpha ,\square \bar \chi^{\dot \alpha } \ne 0, \alpha =1,2;\dot \alpha =\dot 1 ,\dot 2  $ there is no overlap between two (or three) sectors, chiral, antichiral and transversal. We will pay attention to satisfy this condition. 

\section{More on covariant and invariant derivatives}

In this section we provide explicitly some derivatives of supersymmetric functions (not fields) and prove some formulas which in this case will be needed in the sequel; in particular the so called "transfer rules" \cite{S}.
First we compute
\begin{gather} \nonumber
D_\beta X=\varphi_\beta +\theta^\alpha (2m\epsilon_{\beta \alpha })+\bar \theta_{\dot \alpha }(-v_\beta ^{\dot
\alpha }-i\sigma_\beta^{l\dot \alpha}\partial_l f)+ \\ \nonumber +\bar \theta^2 (\psi_{\beta }
-\frac{i}{2}\sigma_{\beta \dot \alpha }^l \partial_l \bar \chi^{\dot \alpha })+\theta^\alpha \bar \theta^{\dot
\alpha }(2\epsilon_{\alpha \beta }\bar \lambda_{\dot \alpha }-i\sigma_{\beta \dot \alpha }^l \partial_l
\varphi_\alpha )+  \\ +\theta^2 \bar \theta_{\dot \alpha }(-i\sigma_\beta^{l\dot \alpha }\partial_l m)+\bar
\theta^2 \theta^\alpha (2\epsilon_{\beta \alpha }d+\frac{i}{2}\sigma_\beta^{l\dot \alpha }\partial_l v_{\alpha
\dot\alpha })-\frac{i}{2}\theta^2 \bar \theta^2 \sigma_{\beta \dot \alpha }^l \partial_l \bar \lambda^{\dot \alpha }
\end{gather}
\begin{gather} \nonumber
\overline{D_{\beta }X}=\bar \varphi_{\dot \beta }+\theta_{\alpha }(-\bar v_{\dot
\beta }^{\alpha }+i\sigma_{\dot \beta }^{l\alpha }\partial_l \bar f)+\bar \theta^{\dot \alpha }(-2\bar m \epsilon_{\dot \alpha \dot \beta})+ \\ \nonumber +\theta^2 (\bar \psi_{\dot \beta }
+\frac{i}{2}\sigma_{\alpha \dot \beta }^l \partial_l \chi^{\alpha })+\theta^\alpha \bar \theta^{\dot
\alpha }(2\epsilon_{\dot \alpha \dot \beta }\lambda_\alpha +i\sigma_{\alpha \dot \beta }^l \partial_l
\bar \varphi_{\dot \alpha })+  \\ 
+\theta^2 \bar \theta^{\dot \alpha }(-2\epsilon_{\dot \alpha \dot \beta }\bar d-\frac{i}{2}\sigma_{\dot \beta }^{l\alpha }\partial_l \bar v_{\alpha \dot \alpha })+\bar \theta^2 \theta_\alpha (i\sigma_{\dot \beta }^{l\alpha }\partial_l \bar m)+\frac{i}{2}\theta^2 \bar \theta^2 \sigma_{\alpha \dot \beta }^l \partial_l \lambda^\alpha 
\end{gather}
\begin{gather} \nonumber
\bar D_{\dot \beta }X=\bar \chi_{\dot \beta }+\theta^{\alpha }(v_{\alpha \dot
\beta }-i\sigma_{\alpha \dot \beta }^l \partial_l f)+\bar \theta_{\dot \alpha } (2n \delta_{\dot \beta }^{\dot \alpha })+ \\ \nonumber +\theta^2 (\bar \lambda_{\dot \beta }
+\frac{i}{2}\sigma_{\alpha \dot \beta }^l \partial_l \varphi^{\alpha })+\theta^\alpha \bar \theta^{\dot
\alpha }(2\epsilon_{\dot \alpha \dot \beta }\psi_\alpha +i\sigma_{\alpha \dot \beta }^l \partial_l
\bar \chi_{\dot \alpha })+  \\ 
+\theta^2 \bar \theta_{\dot \alpha }(2\delta_{\dot \beta }^{\dot \alpha }d+\frac{i}{2}\sigma_{\dot \beta }^{l\alpha }\partial_l \bar v_{\alpha }^{
\dot \alpha })+\bar \theta^2 \theta^\alpha (-i\sigma_{\alpha \dot \beta }^l\partial_l n)+\frac{i}{2}\theta^2 \bar \theta^2 \sigma_{\alpha \dot \beta }^l \partial_l \psi^\alpha 
\end{gather}
We need also
\begin{gather} \nonumber
D_{\beta }\bar X =-\chi_{\beta }+\theta_{\alpha } (2\bar n \delta_{\beta }^{\alpha })-\theta_{\dot \alpha }(\bar v_{\beta }^{\dot \alpha } +i\sigma_{\beta }^{l\dot \alpha }\partial_l \bar f)+\\ \nonumber +\bar \theta^2 (-\lambda_{\beta }
+\frac{i}{2}\sigma_{\beta \dot \alpha }^l \partial_l \bar \varphi^{\bar \alpha })+\theta^{\alpha } \bar \theta^{\dot
\alpha }(-2\epsilon_{\alpha \beta }\bar \psi_{\dot \alpha }+i\sigma_{\beta \dot \alpha }^l \partial_l
\chi_{\alpha })+ \\ 
+\theta^2 \bar \theta^{\dot \alpha }(i\sigma_{\beta \dot \alpha }^l\partial_l \bar n)-\bar \theta^2 \theta_{\alpha }(2\delta_{\beta }^{\alpha }\bar d-\frac{i}{2}\sigma_{\beta }^{l\dot \alpha }\partial_l \bar v_{\dot \alpha }^{\alpha })+\frac{i}{2}\theta^2 \bar \theta^2 \sigma_{\beta \dot \alpha }^l \partial_l \bar \psi^{\dot \alpha } 
\end{gather}
\begin{gather} \nonumber
\overline{D_{\beta }\bar X}=-\bar \chi_{\dot \beta }+\theta_{\alpha }(-v_{\dot
\beta }^{\alpha }+i\sigma_{\dot \beta }^{l\alpha }\partial_l f)+\bar \theta^{\dot \alpha } (-2n \epsilon_{\dot \alpha \dot \beta })+ \\ \nonumber +\theta^2 (-\bar \lambda_{\dot \beta }
-\frac{i}{2}\sigma_{\alpha \dot \beta }^l \partial_l \varphi^{\alpha })+\theta^{\alpha } \bar \theta^{\dot
\alpha }(-2\epsilon_{\dot \alpha \dot \beta }\psi_{\alpha }-i\sigma_{\alpha \dot \beta }^l \partial_l
\bar \chi_{\dot \alpha })+  \\  
+\theta^2 \bar \theta^{\dot \alpha }(-2\epsilon_{\dot \alpha \dot \beta }\bar d-\frac{i}{2}\sigma_{\dot \beta }^{l\alpha }\partial_l \bar v_{\alpha 
\dot \alpha })+\bar \theta^2 \theta_\alpha (i\sigma_{\dot \beta }^{l\alpha }\partial_l n)-\frac{i}{2}\theta^2 \bar \theta^2 \sigma_{\alpha \dot \beta }^l \partial_l \psi^\alpha ) 
\end{gather}
\begin{gather} \nonumber
\bar D_{\dot \beta }\bar X=-\bar \varphi_{\dot \beta }+\theta^{\alpha }(\bar v_{\alpha \dot
\beta }-i\sigma_{\alpha \dot \beta }^l \partial_l \bar f)+\bar \theta_{\dot \alpha } (2\bar m \delta_{\dot \beta }^{\dot \alpha })+ \\ \nonumber +\theta^2 (-\bar \psi_{\dot \beta }
-\frac{i}{2}\sigma_{\alpha \dot \beta }^l \partial_l \chi^{\alpha })+\theta^\alpha \bar \theta^{\dot
\alpha }(-2\epsilon_{\dot \alpha \dot \beta }\lambda_\alpha -i\sigma_{\alpha \dot \beta }^l \partial_l
\bar \varphi_{\dot \alpha })+  \\ 
+\theta^2 \bar \theta_{\dot \alpha }(2\delta_{\dot \beta }^{\dot \alpha }\bar d+\frac{i}{2}\sigma_{\dot \beta }^{l\alpha }\partial_l \bar v_{\alpha }^
{\dot \alpha })+\bar \theta^2 \bar \theta^\alpha (-i\sigma_{\alpha \dot \beta }^l\partial_l \bar m)-\frac{i}{2}\theta^2 \bar \theta^2 \sigma_{\alpha \dot \beta }^l \partial_l \lambda^\alpha 
\end{gather}
in order to prove by inspection that
\begin{gather}
\overline{D_\beta X}=\mp \bar D_{\dot \beta }\bar X, \overline{D_\beta \bar X}=\mp \bar D_{\dot \beta } X
\end{gather}
where $\mp $ means $-$ for even and $+$ for odd $X$ in Grassmann variables. There are no simple formulas for mixed $X$ i.e. neither even nor odd.
We compute further similar expressions for $Q,\bar Q$. For $Q_\beta ,\bar Q_{\dot \beta }$ we find
\begin{gather} \nonumber
Q_\beta X=\varphi_\beta +\theta^\alpha (2m\epsilon_{\beta \alpha })+\bar \theta_{\dot \alpha }(-v_\beta ^{\dot
\alpha }+i\sigma_\beta^{l\dot \alpha}\partial_l f)+ \\ \nonumber +\bar \theta^2 (\psi_{\beta }
+\frac{i}{2}\sigma_{\beta \dot \alpha }^l \partial_l \bar \chi^{\dot \alpha })+\theta^\alpha \bar \theta^{\dot
\alpha }(2\epsilon_{\alpha \beta }\bar \lambda_{\dot \alpha }+i\sigma_{\beta \dot \alpha }^l \partial_l
\varphi_\alpha )+  \\ +\theta^2 \bar \theta_{\dot \alpha }(i\sigma_\beta^{l\dot \alpha }\partial_l m)+\bar
\theta^2 \theta^\alpha (2\epsilon_{\beta \alpha }d-\frac{i}{2}\sigma_\beta^{l\dot \alpha }\partial_l v_{\alpha
\dot\alpha })+\frac{i}{2}\theta^2 \bar \theta^2 \sigma_{\beta \dot \alpha }^l \partial_l \bar \lambda^{\dot \alpha }
\end{gather} 
and
\begin{gather} \nonumber
\bar Q_{\dot \beta }X=\bar \chi_{\dot \beta }+\theta^{\alpha }(v_{\alpha \dot
\beta }+i\sigma_{\alpha \dot \beta }^l \partial_l f)+\bar \theta_{\dot \alpha } (2n \delta_{\dot \beta }^{\dot \alpha })+ \\ \nonumber +\theta^2 (\bar \lambda_{\dot \beta }
-\frac{i}{2}\sigma_{\beta \dot \alpha }^l \partial_l \bar \varphi^{\alpha })+\theta^\alpha \bar \theta^{\dot
\alpha }(2\epsilon_{\dot \alpha \dot \beta }\psi_\alpha -i\sigma_{\alpha \dot \beta }^l \partial_l
\bar \chi_{\dot \alpha })+  \\ 
+\theta^2 \bar \theta_{\dot \alpha }(2\delta_{\dot \beta }^{\dot \alpha }d-\frac{i}{2}\sigma_{\dot \beta }^{l\alpha }\partial_l \bar v_{\alpha }^{\dot \alpha })+\bar \theta^2 \theta^\alpha (i\sigma_{\alpha \dot \beta }^l\partial_l n)-\frac{i}{2}\theta^2 \bar \theta^2 \sigma_{\alpha \dot \beta }^l \partial_l \bar \psi^\alpha 
\end{gather}
The reader can also compute
\[\overline{Q_\beta X},\bar Q_{\dot \beta }\bar X ,Q_\beta \bar X ,\overline{Q_\beta \bar X } \] 
and verify as above that
\begin{gather}
\overline{Q_\beta X}=\mp \bar Q_{\dot \beta }\bar X, \overline{Q_\beta \bar X}=\mp \bar Q_{\dot \beta } X
\end{gather}
where $\mp $ means $ - $ for $X$ even and $ + $ for $X$ odd. \\ 
We need explicitly for several purposes the quadratic derivatives
\begin{gather}\nonumber
D^2 X=-4m+\bar \theta (-4\bar \lambda -2i\bar \sigma^l \partial_l \varphi )+\bar \theta ^2(-4d +2i\partial _l
v^l-\square f)+ \\ 
+\theta \sigma^l \bar \theta (4i\partial _l m)+\bar \theta^2 \theta
(-2i\sigma^l\partial_l \bar \lambda -\square \varphi )+\theta^2 \bar \theta^2 (-\square m)
\end{gather}
\begin{gather}\nonumber
\bar D^2 X=-4n+\theta (-4\psi -2i\sigma^l \partial_l \bar \chi ) +\theta^2 (-4d -2i\partial _l v^l-\square
f)+ \\  
+\theta \sigma^l \bar \theta (-4i\partial _l n)
 +\theta^2 \bar \theta (-2i\bar \sigma^l \partial _l \psi -\square \bar \chi ) +\theta^2 \bar \theta^2 (-\square n) 
\end{gather}
\begin{gather}\nonumber
D^2 \bar X=-4\bar n +\bar \theta (4\bar \psi +2i\bar
\sigma^l \partial_l \chi )+\bar \theta^2 (-4\bar d +2i\partial _l \bar v^l-\square
\bar f)+ \\
+\theta \sigma^l \bar \theta (4i\partial _l \bar n) 
+\bar \theta^2 \theta (2i\sigma^l \partial_l \bar \psi +\square \chi )+\theta^2 \bar \theta^2 (-\square \bar n) 
\end{gather}
\begin{gather}\nonumber
\bar D^2 \bar X=-4\bar m+\theta (4\lambda +2i\sigma^l \partial_l \bar \varphi )+ \theta ^2(-4\bar d -2i\partial _l \bar v^l-\square \bar f)+ \\ 
+\theta
\sigma^l \bar \theta (-4i\partial _l \bar m) +\theta^2 \bar \theta (2i\bar \sigma^l \partial_l \lambda +\square \bar \varphi )
+\theta^2 \bar \theta^2 (-\square \bar m)
\end{gather}
or in a more suggestive way 
\begin{gather}\nonumber
D^2 X=-4m+\bar \theta \bar \xi+\bar \theta ^2(-4d +2i\partial _l
v^l-\square f)+\theta \sigma^l \bar \theta (4i\partial _l m)+\\ +\bar \theta^2 \theta (\frac{1}{2}i\sigma ^l
\partial_l \bar \xi ) +\theta^2 \bar \theta^2 (-\square m)
\end{gather}
\begin{gather}\nonumber
\bar D^2 X=-4n+\theta \eta +\theta^2 (-4d -2i\partial _l v^l-\square
f)+\theta \sigma^l \bar \theta (-4i\partial _l n)+\\
 +\theta^2 \bar \theta (\frac{1}{2}i\bar \sigma^l \partial_l \eta )+\theta^2 \bar \theta^2 (-\square n) 
\end{gather}
where we used the notations
\begin{gather}
\xi =-4 \lambda -2i\sigma^l \partial_l \bar \varphi ,\quad  \eta=-4\psi -2i\sigma^l \partial_l \bar \chi \\
\bar \xi=-4 \bar \lambda -2i\bar \sigma^l \partial_l \varphi , \quad  \bar \eta =-4\bar \psi -2i\bar
\sigma^l \partial_l \chi 
\end{gather}
or explicitly $\xi_\alpha =-4\lambda_\alpha -2i\sigma_{\alpha \dot \beta }^l \partial_l \bar \varphi^{\dot \beta } $ etc. 
We also can write  
\begin{gather}\nonumber
\bar D^2 \bar X=-4\bar m-\theta \xi +\theta ^2 (-4\bar d -2i\partial _l \bar v^l -\square \bar f )+\theta
\sigma^l \bar \theta (-4i\partial _l \bar m)-\\ -\theta^2 \bar \theta (\frac{i}{2}\bar \sigma ^l \partial_l \xi
) 
+\theta^2 \bar \theta^2 (-\square \bar m) \\ \nonumber
D^2 \bar X=-4\bar n-\bar \theta \bar \eta +\bar \theta^2 (-4\bar d +2i\partial _l \bar v^l-\square
f)+\theta \sigma^l \bar \theta (4i\partial _l \bar n)- \\
-\bar \theta^2 \theta (\frac{1}{2}i\sigma^l \partial_l \bar \eta )+\theta^2 \bar \theta^2 (-\square \bar n) 
\end{gather}
Note that in going from $X$ to $\bar X$ we have to replace $\xi ,\bar \xi $ by $-\bar \eta ,-\eta $.
By inspection we find that
\begin{gather}
\overline{D^2 X}=\bar D^2 \bar X , \overline{\bar D^2 X}=D^2 \bar X 
\end{gather}
Another derivation of this formula is
\begin{gather}\nonumber
\overline{D^2 X}=\overline{(D^\alpha D_\alpha )X}=\overline{D^\alpha (D_\alpha  X)}=\mp \bar D^{\dot \alpha }\overline{D_\alpha X}= \\ \nonumber
=(\mp )(\pm )D^{\dot \alpha }(\bar D_{\dot \alpha }\bar X ) =-D^{\dot \alpha }(\bar D_{\dot \alpha }\bar X ) =\bar D_{\dot \alpha }\bar D^{\dot \alpha }\bar X=\bar D^2 \bar X
\end{gather}
From (5.21) follows that
\begin{gather}
\overline{c X}=a \bar X, \quad \overline{a X}=c\bar X, \quad \overline{(c+a)X}=(c+a)\bar X, \quad \overline{TX}=T\bar X
\end{gather}
i.e. the conjugate of $cX$ as superfunction is the superfunction $a\bar X $ etc. Indeed
\begin{gather}\nonumber
\overline{cX}=\overline{\bar D^2 D^2 X}=\overline{\bar D^2 (D^2 X)}=D^2 \overline{(D^2 X)}=D^2 \bar D^2 \bar X=aX \\ \nonumber
\end{gather}
and similarly for $a$. For $T$ we have
\begin{gather}\nonumber
\overline{TX}=\overline{D^\alpha (\bar D^2 D_\alpha X)}=\mp \bar D^{\dot \alpha }\overline{(\bar D^2 D_\alpha X)}= \\ \nonumber
=\mp \bar D^{\dot \alpha }D^2 \overline{D_\alpha X}=(\mp )(\pm )\bar D^{\dot \alpha }D^2 \bar D_{\dot \alpha }\bar X = \\ \nonumber
=-\bar D^{\dot \alpha }D^2 \bar D_{\dot \alpha } \bar X =\bar D_{\dot \alpha }D^2 \bar D^{\dot \alpha } \bar X =T \bar X
\end{gather}
We remind the reader that  $ \bar c ,\bar a , \bar T ,\bar P_c , \bar P_a ,\bar P_T $ were not defined. The situation will be cleared up later when defining Krein- and Hilbert space operator adjoints. 
We have too
\begin{gather}\nonumber
\overline{P_c X}=P_a \bar X, \quad \overline{P_a X}=P_c \bar X, \quad \overline{(P_c +P_a )X}=(P_c +P_a )\bar X, \\ \overline{P_T X}=P_T \bar X, \quad \overline{P_+ X }=P_- \bar X ,\quad \overline{P_- X =P_+ \bar X },\quad \overline{JX}=J\bar X
\end{gather}
Recall that
$J=P_c +P_a -P_T $.
We also need some more relations involving the covariant derivatives $D,\bar D$ which appear in \cite{S} or are consequences of those.
Let $X,Y$ be supersymmetric functions as above. Then we have
\begin{gather}
D_\alpha (XY)=(D_\alpha X)Y\pm X(D_\alpha Y) \\
\bar D_{\dot \alpha }(XY)=(\bar D_{\dot \alpha }X)Y\pm X(\bar D_{\dot \alpha }Y) \\
Q_\alpha (XY)=(Q_\alpha X)Y\pm X(Q_\alpha Y) \\
\bar Q_{\dot \alpha }(XY)=(\bar Q_{\dot \alpha }X)Y\pm X(\bar Q_{\dot \alpha }Y)
\end{gather}
where $\pm $ means $+$ for $X$ even and $-$ for $X$ odd in the Grassmann variables.
It follows that 
\begin{gather}
D^2 (XY)=D^\alpha D_\alpha (XY)=(D^2 X)Y+X(D^2 Y)\pm 2(D^\alpha X)(D_\alpha Y) \\
\bar D^2 (XY)=D_{\dot \alpha }\bar D^{\dot \alpha }(XY)=(D^2 X)Y+X(D^2 Y)\pm 2(\bar D_{\dot \alpha }X)(\bar D^{\dot \alpha } Y) 
\end{gather}
for $X$ even and odd respectively. \\
Now we introduce some kernel functions together with their derivatives which will be used in the next sections. These are functions of the two variables $z_1 =(x_1 ,\theta_1 ,\bar \theta_1 ) ,z_2 =(x_2 ,\theta_2 ,\bar \theta_2 )$ which are supposed to be Taylor expanded in the components of the variables $\theta_1 ,\bar \theta_1 ,\theta_2 ,\bar \theta_2 $. Let us consider
\begin{gather}
k(z_1-z_2)=\delta^8 (z_1 -z_2 )=\delta^2 (\theta_1 -\theta_2)\delta^2 (\bar \theta_1 -\bar \theta_2 )\delta ^4 (x_1-x_2) \\
K_0 (z_1 -z_2)=\delta^2 (\theta_1 -\theta_2)\delta^2 (\bar \theta_1 -\bar \theta_2 )\Omega (x_1-x_2)
\end{gather}
where 
$\delta^2 (\theta_1-\theta_2)=(\theta_1-\theta_2)^2, \delta^2 (\bar \theta_1-\bar \theta_2)=(\bar
\theta_1-\bar \theta_2)^2 $ are Grassmann $\delta-$ functions 
and 
\begin{gather}
\Omega_\rho (x)=\Omega (x)=\frac{1}{(2\pi )^2}\int e^{ipx}d\rho (p), \quad \bar \Omega (x)=\Omega(-x)
\end{gather}
where $d\rho (p)=\rho (p)dp $ is a positive measure such that the integral (5.32) exists as distribution. We have $ \overline{K_0 (z_1 -z_2 )}=K_0 (z_2 -z_1 ) $.\\
Connected to these kernels we have a set of "transfer rules" which are given below:
\begin{gather}
D_{1\alpha }\delta^8 (z_1 -z_2 )=-D_{2\alpha }\delta^8 (z_1 -z_2 ) \\
\bar D_{1\dot \alpha }\delta^8 (z_1 -z_2 )=-\bar D_{2\dot \alpha }\delta^8 (z_1 -z_2 )
\end{gather}
and
\begin{gather}
D_1^2\delta^8 (z_1 -z_2 )=D_2^2\delta^8 (z_1 -z_2 ) \\
\bar D_1^2\delta^8 (z_1 -z_2 )=\bar D_2^2\delta^8 (z_1 -z_2 )
\end{gather}
where the derivative indices refer to the respective variables. Relations of type (5.33)-(5.36) hold for $K_0 (z_1 -z_2 ) $ instead of $k(z_1 -z_2 )$ too, for instance
\begin{gather}
D_{1\alpha }K_0 (z_1 -z_2 )=-D_{2\alpha }K_0 (z_1 -z_2 ) \\
D_1^2 K_0 (z_1 -z_2 )=D_2^2 K_0 (z_1 -z_2 )
\end{gather}
etc. We can now compute
\begin{gather}\nonumber
\bar D_1^2 D_1^2 K_0 (z_ 1-z_2 )=\bar D_1^2 D_2^2 K_0 (z_ 1-z_2 )= \\ 
=D_2^2 \bar D_1^2 K_0 (z_ 1-z_2 )=D_2^2 \bar D_2^2 K_0 (z_ 1-z_2 )
\end{gather}
because in independent variables $[\bar D_1^2 ,D_2^2 ]=0 $.
By the same reasoning we obtain similar "transfer rules" for $a,T$.
It follows that for $K_0 $ depending on $z_1 -z_2 $ we have
\begin{gather}
c_1 K_0 =a_2K_0, \quad
a_1 K_0 =c_2K_0 \quad
T_1 K_0 =T_2K_0
\end{gather}
Assuming that the measure $d\rho (p)$ satisfies a regularity condition at zero momentum (for instance vanishes in momentum space in a small neighborhood of $p=0$ ) we get
\begin{gather}\nonumber
P_{c1} K_0 =P_{a2}K_0 \quad
P_{a1} K_0 =P_{c2}K_0 \quad
P_{T1} K_0 =P_{T2}K_0 \\
J_1 K_0 =J_2 K_0
\end{gather}
Transfer rules holds even for $Q,\bar Q$. We will use only
\begin{gather}
Q_{1\alpha }K_0 (z_1 -z_2 )=-Q_{2\alpha }K_0 (z_1 -z_2 ) \\
Q_1^2 K_0 (z_1 -z_2 )=Q_2^2 K_0 (z_1 -z_2 )
\end{gather}
and similar relations for $\bar Q$.

\section{Some supersymmetric integrals}

In this section we present some results concerning Grassmann (Berezin) integration which will be used in the next sections. In particular we concentrate on integration (including partial integration) of some conjugated (complex and Grassmann) supersymmetric functions (not fields). Recall first the standard notations concerning Berezin integration in supersymmetric context \cite{S,WB}:
\begin{gather} \nonumber
d^2 \theta =\frac{1}{2}d\theta^1 d\theta^2 =-\frac{1}{4}(d\theta^\alpha )(d\theta_\alpha ),  d^2 \bar \theta =-\frac{1}{2}d\bar \theta^1 d\bar \theta^2 =-\frac{1}{4}(d\bar \theta_{\dot \alpha } )(d\bar \theta^{\dot \alpha })  \\ \nonumber
\int d^2 \theta (\theta^2 )=\int d^2 \theta (\theta_\alpha \theta^\alpha )=\int d^2 \theta (-2\theta^1 \theta^2 )=1 \\ \nonumber
\int d^2 \bar \theta^2 (\bar \theta^2 )=1
\end{gather}         
with all other integrals vanishing.
In fact integration coincides with differentiation:
\begin{gather} \nonumber
\int d^2 \theta =\frac{1}{2} \frac{\partial }{\partial \theta^1 }\frac{\partial }{\partial \theta^2 }=\frac{1}{4}\partial^\alpha \partial_\alpha =\frac{1}{4}\partial^2 , \quad
\int d\bar \theta^2 =\bar \partial_{\dot \alpha }\bar \partial^{\dot \alpha }=\frac{1}{4}\bar \partial^2 
\end{gather}
consistent with the definitions above because
\[ \partial^2 \theta^2 =\bar \partial^2 \bar \theta^2 =4 \]
We have $\delta $-function relations, for example
\begin{gather}\nonumber
\int d^2 \theta \delta^2 (\theta^{'}-\theta )f(\theta^1 , \theta^2 , \bar \theta^{\dot 1 }, \bar \theta^{\dot 2 })=f(\theta^{1'}, \theta^{2'},\bar \theta^{\dot 1 },\bar \theta^{\dot 2 } ) \\ \nonumber
\end{gather}
Denoting $d^8 z=d^4 x d^4 \theta ,d^4 \theta =d^2 \theta d^2 \bar \theta $ we have
\begin{gather}
\int d^8 zD_\alpha X=\int d^8 z \bar D_{\dot \alpha }X=0
\end{gather}
for an arbitrary regular function $X$ going to zero at space-time infinity.
From (5.24) and
\begin{gather}
\int d^8 zD_\alpha (XY)=\int d^8 z\bar D_{\dot \alpha }(XY)=0
\end{gather}
it follows that for $X,Y$ going to zero at infinity
\begin{gather}
\int d^8 z XD_\alpha Y=\mp \int d^8 z(D_\alpha X)Y \\ 
\int d^8 z X\bar D_{\dot \alpha }Y=\mp \int d^8 z(\bar D_{\dot \alpha }X)Y 
\end{gather}
according as X is even or odd in the Grassmann variables. There are no simple formulas of type (6.3),(6.4) for $X$ being neither even nor odd. For arbitrary $X$ we have
\begin{gather}
\int d^8 z XD^2 Y=\int d^8 z(D^2 X)Y \\ 
\int d^8 z X\bar D^2 Y=\int d^8 z(\bar D^2 X)Y \\ 
\int d^8 z XP_c Y=\int d^8 z (P_a X)Y \\ 
\int d^8 z XP_a Y=\int d^8 z (P_c X)Y \\ 
\int d^8 z XP_T Y=\int d^8 z (P_T X)Y
\end{gather}
We used $D^2 =D^\alpha D_\alpha =-D_\alpha D^\alpha $ etc. \\ 
Now we state a partial integration result which involves conjugated functions (complex and Grassmann) in the integrands and will be particularly useful for this paper. Indeed from (see (6.3),(6.4))
\begin{gather}\nonumber 
\int d^8 z \bar XD_\alpha Y=\mp \int d^8 z(D_\alpha \bar X)Y \\ \nonumber
\int d^8 z \bar X\bar D_{\dot \alpha }Y=\mp \int d^8 z(\bar D_{\dot \alpha }\bar X)Y 
\end{gather}
and from (5.7) we obtain
\begin{gather} 
\int d^8 z \bar XD_\alpha Y=(\mp )(\mp )\int d^8 z\overline{(\bar D_{\dot \alpha }X)}Y=\int d^8 z\overline{(\bar D_{\dot \alpha }X)}Y \\
\int d^8 z \bar X\bar D_{\dot \alpha }Y=(\mp )(\mp )\int d^8 z\overline{(D_{\alpha }X)}Y=\int d^8 z\overline{(D_{\alpha }X)}Y 
\end{gather}
for $X,Y$ arbitrary satisfying the regularity conditions. \\
Similar relations hold for supersymmetry generators $Q,\bar Q$:
\begin{gather} 
\int d^8 z \bar XQ_\alpha Y=\int d^8 z\overline{(\bar Q_{\dot \alpha }X)}Y \\
\int d^8 z \bar X\bar Q_{\dot \alpha }Y=\int d^8 z\overline{(Q_{\alpha }X)}Y 
\end{gather}
They will be important for what follows. The key for the validity of (6.10)-(6.13) goes back to (3.9) for functions instead of (3.10) for fields. Certainly we have
\begin{gather}
\int d^8 z \bar XD^2 Y=\int d^8 z(D^2 \bar X )Y=\int d^8 z \overline{(\bar D^2 X)}Y  \\ 
\int d^8 z \bar X\bar D^2 Y=\int d^8 z(\bar D^2 \bar X)Y=\int d^8 z \overline{(D^2 X)}Y 
\end{gather}
and
\begin{gather}
\int d^8 z \bar X P_c Y =\int d^8 z (P_a \bar X )Y =\int d^8 z \overline{(P_c X )}Y  \\
\int d^8 z \bar X P_ a Y =\int d^8 z (P_c \bar X )Y =\int d^8 z \overline{(P_a X )}Y\\
\int d^8 z \bar X P_T Y=\int d^8 z (P_T \bar X )Y =\int d^8 z \overline{(P_T X )}Y
\end{gather}
We mention the relations
\begin{gather}
\overline{\int d^8 z X}=\int d^8 z \bar X \\
\overline{\int d^8 z XY}=\int d^8 z \overline{XY}=\int d^8 z\bar Y\bar X
\end{gather}
where on the l.h.s. the bar means numerical complex conjugation whereas on the r.h.s. it stays for complex as well as Grassmann conjugation. We also have
\begin{gather}
\int d^8 z X_i Y_i =0, \quad i=c,a
\end{gather}
Indeed for example
\begin{gather}\nonumber
\int d^8 z X_c Y_c =\int d^8 z X_c P_c Y=\int d^8 z (P_a X_c )Y =0 
\end{gather}
and
\begin{gather}\nonumber
\int d^8 z X_a Y_a =\int d^8 z X_a P_a Y=\int d^8 z (P_c X_a )Y =0 
\end{gather}
There is no similar relation for $P_T $. \\
Although promising, the relations (6.10)-(6.13) and (6.14)-(6.18) unfortunately do not say anything about Hilbert space operator adjointness properties. The reason is that the integrals in superspace which appear in these relations cannot be simply turned into  positive definite sesquilinear forms. The solution to this problem starts in the next section. \\
Before ending let us remark that all considerations in the previous sections concerning functions (not fields) of one supervariable can be generalized to functions of several supervariables. This is not entirely trivial (see \cite{S} for conventions regarding Grassmann integration in several variables). In particular the validity of the relations (6.10)-(6.13) and (6.14)-(6.18) for $X,Y$ depending on the integration variable $z_1 $ and on parameters $ z_2 ,z_3,...$ has to be questioned. The reason is that the Grassmann differentiation and conjugation must respect order. This makes no problem as the reader can easily convince himself.  \\ At the end of presenting all the preparatory material of Sections 1 to 6 the reader might ask himself why we, at extra cost, have abolished fields in favor of functions. The point is that in the next sections we want to do not only algebra, but come across questions touching positivity, scalar products, unitarity etc. for which (wave) functions instead of fields are unavoidable.
 
\section{Indefinite metric: the facts}

In the vector space of supersymmetric functions we want to define positive sesquilinear forms. This is a nontrivial task as experience with integration over Grassmann variables (Berezin integration) shows. Indeed if we form
\begin{gather}
<X,Y>_0 = \int d^8 z \bar X(z)Y(z)
\end{gather}
where $d^8 z =d^4 x d^2 \theta d^2 \bar \theta $, it is easy to see that it is highly indefinite. Nevertheless in the Grassmann sector alone there exist simple examples of positive sesquilinear forms (see for instance \cite{RY}). If we want to cope with the canonical formalism in the Hamiltonian approach to supersymmetric quantum field theory or to other more rigorous approaches than path integrals, we have to start finding positive sesquililear forms of type (7.1). First we write down another form of (7.1). Let $k(z_1 -z_2 )$ be defined as above (see (5.30)).
Then (7.1) will be
\begin{gather}
<X,Y>_k = \int d^8 z_1 d^8 z_2 \bar X(z_1 )k(z_1 -z_2 )Y(z_2 )
\end{gather}
Preparing the way into relativistic superspace we modify (7.2) further to
\begin{gather}
<X,Y> = \int d^8 z_1 d^8 z_2 \bar X(z_1 )K_0 (z_1 -z_2 )Y(z_2 )
\end{gather}
where as in Section 5
\[K_0 (z_1 ,z_2)=K_0 (z_1 -z_2 )=\\ \nonumber
\delta^2 (\theta_1 -\theta_2 )\delta^2 (\bar \theta_1 -\bar \theta_2 )\Omega (x_1 -x_2 )\]
and  
\[\Omega_{\rho}(x)=\Omega(x)=\frac{1}{(2\pi )^2 }\int e^{ipx}d\rho (p), \quad \bar \Omega(x)=\Omega (-x)\]
i.e. $\Omega (x)$ is the inverse Fourier transform of the measure $d\rho (p)=\rho (p)dp $ (for the definition of the Fourier transform see (8.11)).
Although not yet necessary, for application purposes we will assume that the (spectral) measure $d\rho (p)$ is concentrated inside the interior of the forward light cone and eventually that it is Lorentz invariant. The prototype of such a measure is $d\rho (p)=\theta (p_0 )\delta (p^2 +m^2 )dp$ where $m>0$ is the mass, $\delta (p^2 +m^2 )$ the delta-function concentrated on the mass shell $p^2 =-m^2 $ and $ \theta (p_0 ) $ the Heaviside function equal to $ 1$ for positive and to $ 0 $ for negative $p_0 $. The massless case $m=0$ has to be discussed separately (see \cite{C1}).
Experience with quantum field theory suggests that the form $<X,Y>$ should be a good candidate for the (supersymmetric invariant) sesquilinear form defining the physical Hilbert space. Indeed the relation $\overline{<Y,X>}=<X,Y>$ follows from $\overline{K(z)}=K(-z)$. But unfortunately it can be verified that this sesquilinear form is still highly indefinite. Using $P_c +P_a +P_T =1$ it can be written equivalently
\begin{gather}
<X,Y> = \int d^8 z_1 d^8 z_2 \bar X(z_1 )[(P_c +P_a +P_T)K_0 (z_1 -z_2 )]Y(z_2 )
\end{gather}
where $P_c ,P_a ,P_T $ act on the first variable $z_1 $ in $K_0 (z_1 -z_2 )$.
Admitting that $P_c ,P_a ,P_T $ can be hopefully realized as true orthogonal projection operators the indefiniteness of (7.3) or (7.4) seems to be a bad signal: it means that the Hilbert space we are looking for cannot be a direct sum
\begin{gather}
H=H_c \oplus H_a \oplus H_T =H_{c+a }\oplus H_T
\end{gather}
of Hilbert spaces $H_i ,i=c,a,T $ of the chiral, antichiral and transversal sectors in the space of supersymmetric functions. We must conclude that such decompositions which do appear in the physical literature on supersymmetry can be at most formal. In fact this formal decomposition was well-known from the first days of supersymmetry (see for instance the historical review \cite{I}). It resembles the decomposition in electromagnetism into transversal and longitudinal components but this is not quite true; see the discussion in Section 10 for precise statements. In supersymmetry this fact was not taken up seriously at the level of quantization. The reason is that quantization in supersymmetry is generally done by the path integral method which although being extremely successful lies outside Hilbert space and is not able to catch positivity. In electromagnetism it is very much related to the Gupta-Bleuler and St\"uckelberg quantization method. Now, from rigorous point of view, the longitudinal/transversal decomposition in electrodynamics gives rise to indefinite metric (in form of a Krein space) from which the physical Hilbert space can be recovered by a simple procedure \cite{StW,St}. It is reasonable to ask ourself to what extent the supersymmetry induces a similar structure, i.e. to what extent the formal decomposition (7.5) should be replaced by a hopefully rigorous counterpart, for instance
\begin{gather}
H=H_c \oplus H_a \ominus H_T=H_{c+a } \ominus H_T
\end{gather}
with positive scalar product given by
\begin{gather}\nonumber
(X,Y) = \int d^8 z_1 d^8 z_2 \bar X(z_1 )K(z_1 ,z_2 )Y(z_2 )=\\ \nonumber
=\int d^8 z_1 d^8 z_2 \bar X(z_1 )[JK_0 (z_1 -z_2 )]Y(z_2 )=\\
=\int d^8 z_1 d^8 z_2 \bar X(z_1 )[(P_c +P_a -P_T)K_0 (z_1 -z_2 )]Y(z_2 )
\end{gather}
instead of (7.3),(7.4). 
Here 
\begin{gather}
K(z_1 ,z_2 )=JK_0 (z_1 -z_2 )=(P_c +P_a -P_T )K_0 (z_1 -z_2 )    
\end{gather}
is the most important kernel in this paper. The answer to this question is affirmative. Proofs will be provided in the next sections. The integrals in (7.7) can accommodate in momentum space the inverse d'Alembertian if the measure density $\rho (p) $ is concentrated inside the forward light cone. The kernel $K(z_1 ,z_2 )$ is no longer translation invariant (with respect to the Grassmann translations).
As above the operators $P_i ,i=c,a,T $ act on the first variable $z_1 $ of $K_0 $ but they can be transferred to the second variable $z_2 $ of $K_0 $ using (5.41). Because the scalar product (7.7) doesn`t change by this transfer we take the liberty of omitting the hint on which variable they act. Letting $J$ act on the second variable we can write equivalently
\begin{gather}
(X,Y)=\int d^8 z_1 d^8 z_2 \bar X(z_1 )K_0 (z_1 -z_2 )JY(z_2 )
\end{gather}
We denote $P_{c+a }=P_c +P_a $. It follows that
\begin{gather}
(X,Y)=<X,(P_{c+a }-P_T )Y >=<X,JY>
\end{gather}
where both inner products $<.,.>,(.,.)$ are supersymmetric invariant. We have used (6.16-6.18). This is typical for a Krein space and its Hilbert space associate. For precise definitions see Section 10 below. \\
The quest of an indefinite metric inducing the physical Hilbert space in supersymmetry was asked and answered affirmatively in \cite{C1}. Recognizing the Hilbert space of supersymmetry as being generated by the indefinite metric may have applications to rigorous supersymmetric quantum field theory outside path integrals which includes supersymmetric canonical quantization \cite{C2}. \\
There are several proofs of (7.6),(7.7), some of which were sketched in \cite{C1}. In this paper we provide a simple proof which gives not too much insight into the matter and a second one, computationally more involved, worked out in every detail, which provides much more information then the first proof. \\
At this stage a word of caution is necessary:
talking about physical Hilbert space we certainly do not mean the formidable physical Hilbert space of an interacting quantum field theory. From the physical point of view our construction can reach (beside the free field and variants of it as for instance Wick products of a free field or a generalized free field) at most the 2-particle (two point) function of an interacting quantum field as this is illustrated in the last section. From the mathematical point of view we are satisfied by the fact that in our construction the supersymmetric Hilbert space is realized on supersymmetric functions of space-time and Grassmann variables. This might have some advantages (also of physical nature) which will not be described in this paper. \\  
Before starting work let us remark that our statements apply to the relativistic case. We do not touch the supersymmetric quantum mechanics simply because our methods do not apply in this case. In rigorous supersymmetry, as this appears for instance in \cite{V}, the Hilbert space of supersymmetry (relativistic or not) is derived from a general super Hilbert space (which is not a Hilbert space). The indefiniteness is much more stringent because a super Hilbert space contains vectors of imaginary lengths. The two structures: super Hilbert space and our Krein-Hilbert structure are different. In the relativistic case we prefer our structure for reasons to be explained later. It is also interesting to remark that the study of dynamical supersymmetric systems related to the usual BRST quantization \cite{H} also provides hints of indefinite metric. \\
In the next two sections we give proofs of the following statements:
\begin{gather}
(X,Y)=\overline{(Y,X)}\\
(X,X)\ge 0
\end{gather}
for arbitrary supersymmetric $X,Y$ where the bar on the r.h.s. of (7.11) means numerical complex conjugation.
 
\section{Indefinite metric: first proof}

Our first proof doesn't give full insight into the indefinite metric of the $N=1$ superspace but it has the advantage of being computationally simple.
Using the definition of the product $(.,.)$ in (7.7) and (6.19),(6.20) we write
\begin{gather}
(Y,X)=\int d^8 z_1 d^8 z_2 \bar Y(z_1 )JK_0 X(z_2 ) \\
\overline{(Y,X)}=\int d^8 z_1 \bar Z(z_1 )Y(z_1 )
\end{gather}
where 
\begin{gather}
Z(z_1 )=\int d^8 z_2 JK_0 (z_1 -z_2 )X(z_2 )
\end{gather}
We used here the fact that for arbitrary superfunctions $F,G$ (of one or several variables) $\overline{FG}=\bar G \bar F $ holds. Using $ \overline{JK_0 }=J\bar K_0 $ and $\overline{K_0 (z_1 -z_2 )}=K_0 (z_2 -z_1 ) $
we get
\begin{gather}\nonumber
\bar Z (z_1 )=\int d^8 z_2 \bar X (z_2 ) JK_0 (z_2 -z_1 )  
\end{gather}
and
\begin{gather}\nonumber
\overline{(Y,X)}=\int d^8 z_1 [\int d^8 z_2 \bar X(z_2 )JK_0 (z_2 -z_1 )]Y(z_1 )= \\
=\int d^8 z_1 d^8 z_2 \bar X (z_1 )JK_0 (z_1 -z_2 )Y(z_2 )=(X,Y)
\end{gather}
This proves (7.11). With a little more effort the reader can prove that (7.11) remains true even if $J$ is replaced by one of the operators $P_c +P_a ,P_T ,P_+ +P_-  $ or combinations of them with real coefficients.\\
Now we go over to (7.9) taking $Y=X$ and write, using the projection property, transfer rules and partial integration (6.16)-(6.18)
\begin{gather}\nonumber
(X,X)=\int d^8z_1 d^8 z_2 \bar X(z_1 ) (P_c +P_a -P_T )K_0 X(z_2 )=\\ \nonumber 
=\int d^8 z_1 d^8 z_2 \bar X(z_1 ) (P_c ^2 +P_a ^2 -P_T ^2 )K_0 X(z_2 )=\\ \nonumber
=\int d^8 z_1 d^8 z_2 [\sum_{i=c,a }(\overline{P_i X(z_1 )}) K_0 (P_i X(z_2 ))-\overline{P_T X(z_1 }) K_0 (P_T X(z_2 ))]=  \\ 
=\int d^4 x_1 d^4 x_2 \Omega (x_1 -x_2 )(I_c (x_1 ,x_2 )+I_a(x_1 ,x_2 )-I_T (x_1 ,x_2 ))
\end{gather}
with
\begin{gather}\nonumber
I_i (x_1 ,x_2 )=\int d^4\theta_1 d^4 \bar \theta_2 \delta^2 (\theta_1 -\theta_2 )\delta^2 (\bar \theta_1 -\bar \theta_2 )\bar X (z_1 )P_i X(z_2 )= \\
=\int d^4 \theta \bar X_i (x_1 ,\theta ,\bar \theta )X_i (x_2 ,\theta ,\bar \theta ), \quad i=c,a,T
\end{gather}
where $d^4 \theta =d^2 \theta d^2 \bar \theta $. Here $X_i =P_i X , i=c,a,T $ are chiral, antichiral and transversal respectively. We have $\bar X_i =\overline{P_i X }, i=c,a,T $.
We start now the separate study of 
\begin{gather}
(X,X)_i =(X_i ,X_i )=\int d^4 x_1 d^4 x_2  \Omega (x_1 -x_2 )I_i (x_1 ,x_2 ) \quad ,i=c,a,T 
\end{gather}
In the chiral case it follows from Section 4, (4.16) that there are functions  $f,\varphi ,m,v_l =i\partial_l f ,\bar \lambda =\frac{i}{2}\bar \sigma^l \partial_l \varphi  ,d=\frac{1}{4}\square f  $ (other then those which appear in $X, \bar X $, (2.1) and (3.9); this makes the difference to the second proof to follow in Section 9) such that
\begin{gather}
X_c = f+\theta \varphi +\theta^2 m+\theta \sigma^l \bar \theta v_l +\theta^2 \bar \theta \bar \lambda +\theta^2 \bar \theta^2 d
\end{gather}
and therefore
\begin{gather}
\bar X_c = \bar f-\bar \theta \bar \varphi +\bar \theta^2 \bar m+\theta \sigma^l \bar \theta \bar v_l -\bar \theta^2  \theta \lambda +\theta^2 \bar \theta^2 \bar d
\end{gather}
Recall that $\bar \lambda =\frac{i}{2}\bar \sigma^l \partial_l \varphi $ (equivalent to $ \lambda =\frac{i}{2}\sigma^l \partial_l \bar \varphi $) means $\bar \lambda^{\dot \alpha } =\frac{i}{2}\bar \sigma^{l \dot \alpha \beta}\partial_l \varphi_{\beta }$ (equivalent to  $\lambda_{\alpha }  =\frac{i}{2}\sigma_{\alpha \dot \beta }^l \partial_l \bar \varphi^{\dot \beta } $ ).
We find
\begin{gather}\nonumber
I_c (x_1 ,x_2 )=\bar d(x_1 )f(x_2 )-\frac{1}{2}\lambda (x_1 )\varphi (x_2 )+\frac{1}{2}\eta^{lm}\bar v_l (x_1)v_m (x_2 )+ \\
+\bar m(x_1)m(x_2 )-\frac{1}{2}\bar \varphi (x_1 )\bar \lambda (x_2 )+\bar f(x_1 )d(x_2 )  
\end{gather} 
Now we go to the Fourier momentum space. The Fourier transform is defined to be
\begin{gather}
\tilde f (p)=\frac{1}{(2\pi)^2 }\int e^{-ixp}f(x)dx, \quad  f (x)=\frac{1}{(2\pi)^2 }\int e^{ipx}\tilde f(p)dp
\end{gather}
where $xp=x.p$ is the Minkowski scalar product. The derivative $\partial_l $ goes in momentum space as usual to $-\frac{1}{i}p_l $. The following formulas will be used 
\begin{gather}
\int d^4 x_1 d^4 x_2 \bar F(x_1 )\Omega (x_1 -x_2 )G (x_2 )=
\int d^4 p \bar{\tilde F}(p)\rho (p)\tilde G (p)\\
\int d^4 x_1 d^4 x_2 \bar F(x_1 )\Omega (x_1 -x_2 )\square G (x_2 )=
\int d^4 p \bar {\tilde F}(p)\rho (p)(-p^2 )\tilde G (p)
\end{gather}
with the bar being the complex conjugation. We need the case $F=G$. The contributions of $\bar m(x_1 )m(x_2 )$ and of
\begin{gather}\nonumber
\bar d(x_1 )f(x_2 ), \quad
\frac{1}{2}\eta^{lm}\bar v_l (x_1)v_m (x_2 ), \quad
\bar f(x_1 )d(x_2 )  
\end{gather}
in $(X,X)_c $ evaluated with (4.16),(8.12),(8.13) in momentum space are positive. Now we pass to the contributions in $(X,X)_c $ induced by $ -\frac{1}{2}\lambda (x_1 )\varphi (x_2 ) $ and $-\frac{1}{2}\bar \varphi (x_1 )\bar \lambda (x_2 ) $. 
Using (4.16) and
\[ \partial_{1l }\Omega (x_1 -x_2 )=-\partial_{2l}\Omega (x_1 -x_2 ) \]
it is easy to see that they are equal such that it is enough to study
\begin{gather}
A=-\int d^4 x_1 d^4 x_2 \Omega (x_1 -x_2 )\frac{i}{2}\bar \varphi_{\dot \alpha }(x_1 )\bar \sigma^{l\dot \alpha \beta }\partial_l \varphi_{\beta }(x_2 )
\end{gather} 
Indeed we have
\begin{gather}\nonumber
-\frac{1}{2}\lambda (x_1 )\varphi (x_2 )=\frac{i}{4}\partial_l \bar \varphi (x_1 )\bar \sigma^l \varphi(x_2 )=-\frac{i}{4}\bar \varphi (x_1 )\bar \sigma^l \partial_l \varphi(x_2 )\\ \nonumber
-\frac{1}{2}\bar \varphi (x_1 )\bar \lambda (x_2 )=-\frac{i}{4}\bar \varphi (x_1 )\bar \sigma^l \partial_l \varphi (x_2 )
\end{gather}
In order to pass with $A$ to momentum space we need the following variant of (8.12),(8.13)
\begin{gather}
\int d^4 x_1 d^4 x_2 \bar F(x_1 )\Omega (x_1 -x_2 )(-i\frac{\partial }{\partial x_2^l })H(x_2 )=
\int d^4 p \bar{\tilde F}(p)\rho (p)p_l \tilde H (p)
\end{gather}
which in a matrix generalization reads
\begin{gather}\nonumber
\int d^4 x_1 d^4 x_2 \Omega (x_1 -x_2 )\bar F (x_1 )M(-i\frac{\partial }{\partial x_2 })H(x_2 )= \\
=\int d^4 p \rho (p)\bar{\tilde F}(p)M(p)\tilde H (p)
\end{gather}
where $F,H$ are vectors and $M$ a matrix with entries depending on $-i\frac{\partial }{\partial x_2 }=(-i\frac{\partial }{\partial x_2^l }) $. Using (8.16) with $F=H$ we obtain in momentum space
\begin{gather}
A=\frac{1}{2}\int d^4 p \rho (p) \bar {\tilde \varphi }_{\dot \alpha }(p)\bar \sigma^{l\dot \alpha \beta }p_l \tilde{\varphi }_{\beta }(p)=\frac{1}{2}\int d^4 p \rho (p)\bar {\tilde \varphi }(p)(\bar \sigma p)\tilde{\varphi }(p) 
\end{gather}
and this is positive because the matrix $\bar \sigma p=\bar \sigma^l p_l $ (as well as $ \sigma p $) is positive definite in the forward light cone where the measure $d\rho (p) $ is concentrated. Certainly we were carefully enough in order to have at this final stage of computation standard index positions in the van der Waerden $\sigma ,\bar \sigma $ (matrix interpretation). The positivity of the matrix $\bar \sigma p $ (and $ \sigma p $) can be easily verified by reading up its trace and determinant. We remind that our convention is $\sigma^0 =1 $. \\
The computation of the antichiral contribution to (8.5) is similar and gives a positive result too.  \\
The transversal contribution to (8.5) is more interesting because unexpected. Although it looks similar to the other two contributions it turns out to be negative! Indeed we have (with other coefficients than those which appear in $X,\bar X $)
\begin{gather}\nonumber
X_T = f+\theta \varphi +\bar \theta \bar \chi +\theta \sigma^l \bar \theta v_l +\theta^2 \bar \theta \bar \lambda +\bar \theta^2 \theta \psi +\theta^2 \bar \theta^2 d  \\ \nonumber
\bar X_T = \bar f-\theta \chi -\bar \theta \bar \varphi +\theta \sigma^l \bar \theta \bar v_l -\theta^2 \bar \theta \bar \psi -\bar \theta^2 \theta \lambda +\theta^2 \bar \theta^2 \bar d 
\end{gather}
with (4.18)
\[ \partial_l v^l =0, \bar \lambda =-\frac{i}{2}\bar \sigma^l \partial_l \varphi ,\psi  =-\frac{i}{2}\sigma^l \partial_l \bar
\chi  , d=-\frac{1}{4}\square f \]
and hence
\begin{gather}\nonumber
I_T (x_1 ,x_2 )=\bar d(x_1 )f(x_2 )-\frac{1}{2}\lambda (x_1 )\varphi (x_2 )-\frac{1}{2}\bar \psi (x_1 )\bar \chi (x_2 )- \\ -\frac{1}{2}\eta^{lm}\bar v_l (x_1)v_m (x_2 ) 
-\frac{1}{2}\bar \varphi (x_1 )\bar \lambda (x_2 )-\frac{1}{2}\chi (x_1 )\psi (x_2 )+\bar f(x_1 )d(x_2 )  
\end{gather} 
with $\partial_l v^l =0 $. The only new term to be studied is 
\[ -\frac{1}{2}\eta^{lm}\bar v_l (x_1)v_m (x_2 )=-\frac{1}{2}\bar v_l (x_1)v^l (x_2 ),\quad \partial_l v^l =0 \]
It has to be subtracted in (8.5) such that we have to prove that $ \bar v_l (x_1 )v^l (x_2 ) $ gives a positive contribution. For proving this assertion we use a nice old argument. First note that in momentum space we have $p_l \tilde v^l (p)=0 $. Suppose that $\tilde v (p)$ has real components $\tilde v^l (p) $. Then the relation $ p_l \tilde v^l (p)=0 $ means that the vector with components $\tilde v^l (p)$ is orthogonal in euclidean sense to the vector $p=(p_l ) $. But the vector $p$ is confined to the interior of the light cone because the measure $d\rho (p)$ is and therefore the vector with components $\tilde v^l (p)$ is space-like. This means that in the metric $(-1,1,1,1)$ we have $\tilde v_l (p)\tilde v^l (p)>0 $ and we obtain the desired result. If $\tilde v (p)$ is complex i.e. some of its components or all of them are, then we split it in a real and an imaginary part and apply twice the same argument to prove that $\overline {\tilde v_l }(p)\tilde v^l (p)>0 $. \\
By this the first proof of the indefinite metric (Krein-Hilbert structure) of the $N=1$ superspace is completed.

\section{Indefinite metric: second proof}

In this section we use $P_T =1-P_c -P_a $ and explicitly compute
\begin{gather}\nonumber
(X,X)=\int d^8 z_1 d^8 z_2 \bar X(z_1 ) (P_c +P_a -P_T )K_0 X(z_2 )= \\ \nonumber
=(X,X)_c+(X,X)_a-(X,X)_T 
=(X,X)_{a+c}-(X,X)_T = \\ \nonumber
=\int d^8z_1 d^8 z_2 \bar X(z_1 ) (2P_c +2P_a -1)K_0 X(z_2 )= \\ 
=\int d^4 x_1 d^4 x_2 \Omega (x_1 -x_2 )(2I_c (x_1 ,x_2 )+2I_a (x_1 ,x_2 )-I_0 (x_1 ,x_2 ))
\end{gather}
and separately
\begin{gather}\nonumber
(X,X)_{c+a }=\int d^8z_1 d^8 z_2 \bar X(z_1 )(P_c +P_a )K_0 X(z_2 )= \\ \nonumber
=\int d^4 x_1 d^4 x_2 \Omega (x_1 -x_2 )(I_c (x_1 ,x_2 )+I_a (x_1 ,x_2 )) \\ \nonumber
(X,X)_T =\int d^8z_1 d^8 z_2 \bar X(z_1 )P_T K_0 X(z_2 )= \\ \nonumber
=\int d^4 x_1 d^4 x_2 \Omega (x_1 -x_2 )(-I_c (x_1 ,x_2 )-I_a (x_1 ,x_2 )+I_0 (x_1 ,x_2 ))
\end{gather}
in terms of the coefficients of $X$. In (9.1) we denoted 
\begin{gather}
I_c =I_c (x_1 ,x_2 )=\frac{1}{16\square }\int d^2\theta d^2 \bar \theta \bar D^2 \bar X (x_1 ,\theta ,\bar \theta )D^2 X(x_2 ,\theta ,\bar \theta ) \\
I_a =I_a (x_1 ,x_2 )=\frac{1}{16\square }\int d^2 \theta d^2 \bar \theta D^2 \bar X (x_1 ,\theta ,\bar \theta )\bar D^2 X(x_2 ,\theta ,\bar \theta ) 
\end{gather}
and
\begin{gather}
 I_0 =I_0 (x_1 ,x_2 )=\int d^2 \theta d^2 \bar \theta \bar X (x_1 ,\theta ,\bar \theta )X(x_2 ,\theta ,\bar \theta ) 
\end{gather}
Here $I_c ,I_a ,I_0 $ (and later $I_T ,I_{\pm}$) are integrands and this explains the free manipulations with d'Alembertians.
Note that $I_i ,i=c,a,T $ in this section are different from those of Section 8.
We use (5.11)-(5.14) and obtain
\begin{gather}\nonumber
16\square I_c =-4\bar m(x_1 )(-\square m(x_2 ))+ \\ \nonumber
+\frac{1}{2}(4\lambda (x_1 )+2i\sigma^l \partial_l \bar \varphi (x_1 ))(-2i\sigma^m \partial_m \bar \lambda (x_2)- 
\square \varphi (x_2 ))+ \\ \nonumber
+(-4\bar d(x_1)
-2i\partial _l \bar v^l x_1 )-\square \bar f (x_1 ))(-4d(x_2 ) +2i\partial _l v^l (x_2 )-\square f(x_2 ))-  \\ \nonumber
-\frac{1}{2}\eta^{lm}(-4i\partial_l\bar m(x_1))(4i\partial_m m(x_2 ))
+\frac{1}{2}(2i\bar \sigma^l\partial_l \lambda (x_1)+\square \bar \varphi (x_1 ))\times \\
\times (-4\bar \lambda (x_2 )-2i\bar \sigma^l \partial_l \varphi (x_2 ))+(-\square \bar m (x_1 )(-4m_2 (x_2 ))
\end{gather}
\begin{gather}\nonumber
16\square I_a =-4\bar n(x_1 )(-\square n(x_2 ))+ \\ \nonumber
+\frac{1}{2}(4\bar \psi (x_1 )+2i\bar \sigma^l \partial_l \chi (x_1 ))(-2i\bar \sigma^m \partial_m \psi (x_2)- 
\square \bar \chi (x_2 ))+ \\ \nonumber
+(-4\bar d(x_1)
+2i\partial _l \bar v^l x_1 )-\square \bar f (x_1 ))(-4d(x_2 ) -2i\partial _l v^l (x_2 )-\square f(x_2 ))-  \\ \nonumber
-\frac{1}{2}\eta^{lm}(4i\partial_l\bar n(x_1))(-4i\partial_m n(x_2 ))
+\frac{1}{2}(2i\sigma^l\partial_l \bar \psi (x_1)+\square \chi (x_1 ))\times \\ 
\times (-4\psi (x_2 )-2i\sigma^l \partial_l \bar \chi (x_2 ))+(-\square \bar n (x_1 )(-4n_2 (x_2 ))
\end{gather}
Note that in $I_c +I_a $ the mixed contribution of $\bar v$ with $d,f$ and of $v$ with $\bar d ,\bar f $ vanish. Moreover $I_c $ does not depend on $\chi ,\psi $ and $I_a $ does not depend on $\varphi ,\lambda $.\\
We obtain
\begin{gather}\nonumber
I_c +I_a =\bar m(x_1 )m(x_2 )+\bar n(x_1 )n(x_2 )+ \\ \nonumber
+\frac{1}{8\square }(4\bar d (x_1 )+\square \bar f (x_1 ))(4d (x_2 )+\square f (x_2 ))+\frac{1}{2\square }(\partial_l \bar v^l (x_1 ))(\partial_m v^m (x_2 ))+ \\ \nonumber
+\frac{1}{2\square}\lambda (x_1 )\sigma^l (-i\partial_l )\bar \lambda (x_2 )+\frac{1}{8}\bar \varphi (x_1 )\bar \sigma^l (-i\partial_l )\varphi (x_2 )- \\ \nonumber
-\frac{1}{4}\lambda (x_1 )\varphi (x_2 )-\frac{1}{4}\bar \varphi (x_1 )\bar \lambda (x_2 )+ \\ \nonumber 
+\frac{1}{2\square}\bar \psi (x_1 )\bar \sigma^l (-i\partial_l )\psi (x_2 )+\frac{1}{8}\chi (x_1 )\sigma^l (-i\partial_l )\bar \chi (x_2 )- \\ 
-\frac{1}{4}\bar \psi (x_1 )\bar \chi (x_2 )-\frac{1}{4}\chi (x_1 )\psi (x_2 )
\end{gather}
where we have used relations of the type
\begin{gather}\nonumber
\int d^4 x_1 d^4 x_2 (\sigma^l \partial_l \bar \varphi (x_1 ))(\sigma^m \partial_m \bar \lambda (x_2 ))= \\ \nonumber
=-\int d^4 x_1 d^4 x_2 \bar \varphi (x_1 )\square \bar \lambda (x_2 ) \\ \nonumber
\int d^4 x_1 d^4 x_2 (\bar \sigma^l \partial_l \lambda (x_1 ))(\bar \sigma^m \partial_m \varphi (x_2 ))= \\ \nonumber
=-\int d^4 x_1 d^4 x_2 \lambda (x_1 )\square \varphi (x_2 )  
\end{gather}
which can be proved using (3.6),(3.7).
From $2I_c +2I_a $ we subtract
\begin{gather}\nonumber
I_0=\bar f(x_1 )d(x_2 )-\frac{1}{2}\chi (x_1 )\psi (x_2 )-\frac{1}{2}\bar \varphi (x_1 ) \bar \lambda (x_2 )+\bar n(x_1 )n(x_2 )+ \\ \nonumber
+\bar m(x_1 )m(x_2 )-\frac{1}{2}\eta^{lm}\bar v_l (x_1 )v_m (x_2 )-\frac{1}{2}\bar \psi (x_1 )\bar \chi (x_2 )- \\ 
-\frac{1}{2}\lambda (x_1 )\varphi (x_2 )+ \bar d (x_1 )f(x_2 )
\end{gather}
The nice fact is that all mixed contributions of coefficient functions in $(X,X) $ considered as integrands (i.e. taking into account the minus one factor at the transfer of $\partial_l $ from the variable $x_1 $ to the variable $x_2 $ or vice-versa) in $I_c +I_a -I_T =2(I_c +I_a ) -I_0 $ vanish. \\
The computations above are elementary. There are some points which might be mentioned, for instance the contributions in $I_c $ of the type 
\[(2i\bar \sigma^l \partial_l \lambda (x_1 ))(-4\bar \lambda (x_2 ))\]
must be read correctly: 
\begin{gather}\nonumber
-8i(\bar \sigma^l \partial_l )\lambda (x_1 )\bar \lambda (x_2 )=-8i(\bar \sigma^l \partial_l \lambda (x_1 ))_{\dot \alpha }\bar \lambda (x_2 )^{\dot \alpha } 
=8\lambda (x_1 )\sigma^l (-i\partial_l )\bar \lambda (x_2 )
\end{gather} 
i.e. the summation in $\dot \alpha $ goes south-east to north-west (not north-west to south-east which would give the wrong sign), this being clear from the provenience of this term in $I_c $. Having computed $I_c +I_a $ and $I_0 $ we can obtain $(X,X)$ using (9.1).
In order to write down the result let us denote by $\Vert m\Vert^2 $,...the contributions of $ m,\bar m  $,...in $ (X,X)=\Vert X \Vert^2  $ as given below
\begin{gather}\nonumber
\Vert m\Vert^2 =\int d^4 x_1 d^4 x_2 \Omega (x_1 -x_2 )\bar m(x_1 )m(x_2 ) \\ \nonumber
\Vert n\Vert^2 =\int d^4 x_1 d^4 x_2 \Omega (x_1 -x_2 )\bar n(x_1 )n(x_2 ) \\ \nonumber
\Vert f \Vert^2 =\frac{1}{4}\int d^4 x_1 d^4 x_2 \Omega (x_1 -x_2 )\bar f (x_1 )\square f(x_2 ) \\ \nonumber
\Vert d \Vert^2 =4\int d^4 x_1 d^4 x_2 \Omega (x_1 -x_2 )\bar d (x_1 )\frac{1}{\square }d(x_2 ) \\ \nonumber
\Vert v \Vert^2 =\frac{1}{2}\int d^4 x_1 d^4x_2 \Omega (x_1 -x_2 )[\bar \Delta (x_1 )\frac{1}{\square }\Delta(x_2 )+\bar V_l (x_1 )V^l (x_2 )] 
\end{gather}
and \begin{gather} \nonumber
\Vert \varphi \Vert^2 =\frac{1}{4}\int d^4 x_1 d^4x_2 \Omega (x_1 -x_2 )\bar \varphi (x_1 )\bar \sigma^l (-i\partial_l )\varphi (x_2 ) \\ \nonumber
\Vert \lambda \Vert^2 =\int d^4 x_1 d^4x_2 \Omega (x_1 -x_2 )\lambda (x_1 )\frac{1}{\square }\sigma^l (-i\partial_l )\bar \lambda (x_2 ) \\ \nonumber
\Vert \chi \Vert^2  =\frac{1}{4}\int d^4 x_1 d^4x_2 \Omega (x_1 -x_2 )\chi (x_1 ) \sigma^l (-i\partial_l )\bar \chi (x_2 ) \\ \nonumber
\Vert \psi \Vert^2  =\int d^4 x_1 d^4x_2 \Omega (x_1 -x_2 )\bar \psi (x_1 )\frac{1}{\square }\bar \sigma^l (-i\partial_l )\psi (x_2 )  
\end{gather}
where 
\begin{gather}
\Delta (x)=\partial_l v^l (x),\quad V_l (x)=v_l (x)-\partial_l \frac{\partial_m v^m (x)}{\square } 
\end{gather}
Here we applied the relation often used in electrodynamics
\begin{gather}
\frac{1}{\square }(\partial_l \bar v^l (x_1 ))(\partial_m v^m (x_2 ) ))+\bar v_l (x_1 )v^l (x_2 )=\bar V_l (x_1 )V^l (x_2 ) \\ 
\partial_l V^l (x)=0
\end{gather}
The relation (9.11) shows that $V^l (x)$ is space-like in momentum space and the disscusion at the end of Section 8 shows that $\bar V_l (x_1 )V^l (x_2 ) $ gives a positive contribution and therefore $\Vert v\Vert^2 $ is positive.
We get
\begin{gather}\nonumber
\Vert X \Vert^2 =(X,X)=\Vert f \Vert^2 +\Vert \varphi \Vert^2 +\Vert \chi \Vert^2 +\Vert m\Vert^2 +\Vert n\Vert^2
+\Vert v\Vert^2 + \\ \nonumber
+\Vert \lambda \Vert^2 +\Vert \psi \Vert^2 +\Vert d \Vert^2 
\end{gather}
In fact we could have computed
\begin{gather}\nonumber
(X_1 ,X_2 )=(f_1 ,f_2 )+(\varphi_1 ,\varphi_2 ) +(\chi_1 ,\chi_2 )+(m_1 ,m_2 ) +(n_1 ,n_2 )+ \\ 
+(v_1 ,v_2 )+(\lambda_1 ,\lambda_2 )+(\psi_1 , \psi_2 )+(d_1 ,d_2 )  
\end{gather}
where the scalar products $(f_1 ,f_2 )$ etc. can be read up from the corresponding norms. Roughly speaking our Hilbert space turns up to be an orthogonal direct sum
\begin{gather}
H=\oplus H_{\mbox{ components }}
\end{gather}
This is a surprising simple result. Note that the supersymmetry is responsible for the specific numerical factors and d'Alembertians in the norms and scalar products respectively. By this, the second, explicit proof of indefinite metric and of the Hilbert space scalar product generated by it is completed. \\
Analog computations provide the results for $(X,X)_{c+a }, (X,X)_T $. The result for $(X,X)_{c+a}=\Vert X\Vert_{c+a }^2 $ can be written in compact form using $\xi ,\bar \xi $ introduced in Section 5. Because we are not especially interested in this scalar product we will not write it down explicitly. We concentrate on $-(X,X)_T =\Vert X \Vert_T ^2 $ obtaining from $I_T =I_0 -I_c -I_a $ 
\begin{gather}\nonumber
-(X,X)_T =\int d^4 x_1 d^4 x_2 \Omega (x_1 -x_2 )[\frac{1}{8\square }(4\bar d(x_1 )-\square \bar f(x_1 ))(4d(x_1 )-\square f(x_1 ))+ \\ \nonumber
+\frac{1}{2}\bar V_l (x_1 )V^l (x_2 )+ \\ 
+\frac{1}{32\square }\xi^T (x_1 )\sigma^l (-i\partial_l )\bar \xi^T (x_2 )+\frac{1}{32\square }\bar \eta^T (x_1 )\bar \sigma^l (-i\partial_l )\eta^T (x_2 )] 
\end{gather}
where
\begin{gather}
\xi^T =-4 \lambda +2i\sigma^l \partial_l \bar \varphi ,\quad  \eta^T =-4\psi +2i\sigma^l \partial_l \bar \chi \\
\bar \xi^T =-4 \bar \lambda +2i\bar \sigma^l \partial_l \varphi , \quad  \bar \eta^T =-4\bar \psi +2i\bar
\sigma^l \partial_l \chi 
\end{gather}
The scalar product $-(X,Y)_T $ of two supersymmetric functions can be inferred from (9.14). Note that in $\Vert X \Vert_T $ and $ (X,Y)_T $ the "auxiliary functions" $m,n $ do not appear at all. The discussion of results is deferred to the next section.

\section{Indefinite metric: discussion of results}

Let us start by giving the precise definition of a Krein space together with its Hilbert space counterpart. Assume that in a Hilbert space $H$ there is given a self-adjoint operator $J$ satisfying the relation $J^2 =1 $ and introduce the projections $ P_1 =\frac{1}{2}(1+J),P_2 =\frac{1}{2}(1-J) $. The projections $P_1 ,P_2 $ generate a decomposition of the Hilbert space $H$ into the direct sum of two orthogonal subspaces. We have for $ X\in  H $ the unique decomposition $X=P_1 X \oplus P_2 X =X_1 \oplus X_2 $. Introduce in $H$ a new non-degenerate inner product $<.,.>$ such that
\begin{gather}
<X,Y>=(X,JY)
\end{gather}
We have $J=P_1 -P_2 $ such that
\begin{gather}
<X,Y>=(X_1 ,Y_1 )-(X_2 ,Y_2 ) 
\end{gather}
and (for $J\neq \pm 1 $) we generate an indefinite metric in $H$. The space $H$ looked at as a vector space with inner product $<.,.>$ is called a Krein space K. We call the couple of two spaces $ (K,H) $ together with the sesquilinear form $<.,.>$ on $K$ and the scalar product $(.,.)$ on $H$ a Krein-Hilbert or a Hilbert-Krein structure. The terminology is not standard; the reader may reject it. \\
In applications to physics it might happen that we first construct a sesquilinear form $<.,.> $ on a vector space $H$, choose an operator $J$ and verify that $(X,Y)=<X,JY>$ is positive definite. This means that $H$ is actually a Hilbert space. We have only to check that $J$ is Hilbert self-adjoint, $J^2 =1$ and finally $<X,Y>=(X,JY)$. This is the way we constructed our Krein-Hilbert structure which proves indefinite metric in superspace. This indefinite metric of the $N=1 $ supersymmetry is similar to the corresponding structure in electrodynamics. Before explaining the matter we have to add a word of caution. Talking about electrodynamics we mean here massive electrodynamics. Indeed in this paper we are confined to the case in which the defining measure $d\rho (p) $ is supported in momentum space inside the light cone and doesn't touch the boundary. This condition is needed in order to make well-define d'Alembertians in many denominators. It also kills the null-vectors (in our case as well as in the massive electrodynamics too). With some extra work we can show that this condition can be removed at the cost of restricting the allowed supersymmetric test functions (by standard factorization followed by completion). For the convenience of the reader we recall that in electrodynamics the indefiniteness \cite{St} in the case of a vector field appears in form of a Krein space too which at the level of test functions $v=(v_l )$ boils down either to the physical transversal Hilbert space of Gupta-Bleuler (obtained by imposing the above mentioned subsidiary condition) or more generally to the St\"uckelberg Hilbert space 
\begin{gather}
H=H_t \ominus H_l
\end{gather}
where $H_l $ is the longitudinal contribution. It turns out to arise in the process of quantization with wrong sign and therefore has to be subtracted. The "subsidiary condition" annihilates the longitudinal contribution and we would stay with the transversal $H=H_t $.
In (10.3) from technical point of view $H_t ,H_l $ are obtained with the help of projection matrices
\begin{gather}
P_{trans} =P_{trans}^{lm}=\eta^{lm}-\frac{\partial^l \partial^m }{\square } \\
P_{long} =P_{long} ^{lm}=\frac{\partial^l \partial^m }{\square }
\end{gather}
where $l,m=0,1,2,3 $.
The relations (10.4),(10.5) are read as matrices applied to vectors:
\begin{gather}
P_{trans}^{lm}v_m =(\eta^{lm} -\frac{\partial^l \partial^m }{\square })v_m =V^l \\
P_{long}^{lm}v_m =\frac{\partial^l \partial^m }{\square }v_m =\partial^l \frac{\Delta }{\square }
\end{gather}
The transversality (Lorentz) condition is $\partial P_{trans}v=0$ or explicitely
\begin{gather}
\partial_l P_{trans}^{lm}v_m =0
\end{gather}
The selfadjoint $P_{trans} ,P_{long} $ are  projections $P_{long}^2 =P_{long} ,P_{trans}^2 =P_{trans} $ and we have $P_{long} +P_{trans} =1, P_{long} P_{trans} =P_{trans} P_{long} =0 $. Finally the considerations of the preceding section can be used to show that $ H_t \oplus H_l $ is a Krein space and gives indefinite metric whereas the right Hilbert space is $H=H_t \ominus H_l $. The argument showing that $P_{trans} $ produces a positive contribution and $P_{long}$ a negative one is the same as in supersymmetry (see Sections 8,9). \\
Now, what we obtained in supersymmetry 
\begin{gather}\nonumber
H=H_c \oplus H_a \ominus H_T=H_{c+a } \ominus H_T
\end{gather}
is very similar. 
The $v$-norm from the preceding Section 
\begin{gather}\nonumber
\Vert v \Vert^2 =\frac{1}{2}\int d^4 x_1 d^4x_2 \Omega (x_1 -x_2 )[\bar V_l (x_1 )V^l (x_2 )+\bar \Delta (x_1 )\frac{1}{\square }\Delta(x_2 )] 
\end{gather}
shows that the electrodynamic $H_t \ominus H_l $ is contained (but not exhausts) the supersymmetric $H $. The first term with integrand $\bar V_l (x_1 )V^l (x_2 )$ refers to the transversal whereas the second one with integrand $\bar \Delta (x_1 )\frac{1}{\square }\Delta(x_2 )$ (sign changed!) to the longitudinal contribution. 
The conclusion is that the Krein-Hilbert structure of supersymmetry is very similar to the corresponding structure in electrodynamics being overimposed on it. More precisely the Krein transversal/longitudinal structure of electrodinamics is included in the Krein chiral plus antichiral/transversal structure of supersymmetry. \\
Last but not least: the minus sign of the supersymmetric Krein space has nothing to do with the celebrated minus sign of the supertrace. 

\section{Covariant derivative operators and supersymmetric generators}

We have seen in Section 10 that a general Krein-Hilbert structure $ (K,H) $ is given by $<X,Y>=(X,JY),J^2 =1,J=J^{\dagger} $ where the dagger represents the Hilbert space adjoint operator. We define two types of adjoint operator: the Krein adjoint $A^+ $ called also $J$-adjoint and the Hilbert adjoint $A^{\dagger} $ of a given operator $A$. In order to simplify the matter we will leave out the details concerning the domains of definition, existence of adjoints etc. The Krein adjoint is defined through
\begin{gather}
<X,AY>=<A^+ X,Y>
\end{gather}
The relation between the Hilbert space adjoint $A^{\dagger }$ and the Krein space adjoint $A^+ $ of $A$ is
\begin{gather}
A^+ =JA^{\dagger }J, \quad
A^{\dagger }=JA^+ J
\end{gather}
As in the case of Hilbert seftadjointness $A^{\dagger }=A$, $A$ is said to be $J$-self adjoint if $A^+ =A$. Moreover if $[J,A]=0$ then $ A^+ =A^{\dagger } $. \\
After these general statements we come back to our particular Krein-Hilbert structure. Here $J=P_c +P_a -P_T $ and
\begin{gather}\nonumber
<X,JY> = \int d^8 z_1 d^8 z_2 \bar X(z_1 )K_0 (z_1 -z_2 )JY(z_2 )= \\ 
\int d^8 z_1 d^8 z_2 \bar X(z_1 )[JK_0 (z_1 -z_2 )]Y(z_2 ) = (X,Y)
\end{gather}
We consider now the operators $ D_\alpha, \bar D_{\dot \alpha },D^2 ,\bar D^2 ,Q_\alpha ,\bar Q_{\dot \alpha }$ etc. and ask ourself to what extent the bar represents the adjoint and in the affirmative case which adjoint. From the relations proved in Section 6 follows that  $ \bar D_{\dot \alpha }, \bar D^{\dot \alpha }, \bar Q_{\dot \alpha }, \bar Q^{\dot \alpha }$ are $J$-adjoints of $ D_{\alpha }, D^{\alpha }, Q_{\alpha }, Q^{\alpha }$. We conclude that for these operators bar is identical to the $J$-adjoint. What is more interesting is the question concerning the Hilbert adjoints. We start with the covariant derivative. It is not difficult to convince ourself that $D_{\alpha }, D^{\alpha }, \bar D_{\dot \alpha }, \bar D^{\dot \alpha }$ do not commute with $J=P_c +P_a -P_T $. It follows that $\bar D_{\dot \alpha }, \bar D^{\dot \alpha }$ are not Hilbert space adjoints of $D_{\alpha }, D^{\alpha } $. The correct answer is
\begin{gather}
(D_{\alpha })^{\dagger }=J\bar D_{\dot \alpha }J \\ 
(D^{\alpha })^{\dagger }=J\bar D^{\dot \alpha }J
\end{gather}
But we have $(D^2)^{\dagger }=\bar D^2, (\bar D^2)^{\dagger }=D^2  $ and therefore $(P_+ )^{\dagger }=P_- ,(P_- )^{\dagger }=P_+ ,(P_T )^{\dagger}=P_T $. Furthermore $P_c ,P_a ,P_T $ are Hilbert self adjoints; a property which makes them true orthogonal projection operators (with $P_c +P_a +P_T =1$). One has to contrast formulas like
\[\overline{P_c X }=P_a \bar X, \overline{P_a X }=P_c \bar X \]
to
\[(P_c )^{\dagger } =P_c ,(P_a )^{\dagger } =P_a \]
On the contrary it is pleasant to see that the supersymmetric generators make no problems at all because they commute with $D$-operators (and therefore with the $J$ operator). It follows that 
\begin{gather}
\bar Q_{\dot \alpha }=(Q_\alpha )^+ =(Q_\alpha )^{\dagger }
\end{gather} 
Having realized the generators of the translation supergroup (certainly including the translations $P$) as Hilbert space operators with sound adjointnes properties, the first idea we can have is to exponentiate them in order to generate group elements. Formally Salam and Strathdee beautifully 
showed that this exponentiation has to be done using Grassmann parameters $\epsilon ,\bar \epsilon $ ($\epsilon ,\bar \epsilon $ here have nothing to do with the metric tensors in Section 3) in the form $\mbox{ exp }(i\epsilon Q+i\bar \epsilon \bar Q ) $. The problem we encounter in our rigorous framework is that the Grassmann parameters $\epsilon ,\bar \epsilon $ kick us out of the Hilbert space of supersymmetric functions of the variables $x,\theta ,\bar \theta $ on which the operators $Q, \bar Q$ are realized. At first glance this seems to be unpleasant and we have to find a way out. There are several possibilities. One of them is to use Harish-Chandra pairs \cite{V} in order to cope with the representation theory of supergroups. We will not follow this route here but apply ideas of distribution theory in the supersymmetric context i.e. we smear the above exponential by test functions in the parameters $\epsilon ,\bar \epsilon $. A similar procedure was proposed in \cite{O} on the bases of Hopf algebra (group algebra) considerations. A potential application is a rigorous Wigner type theory of unitary irreducible representations of the supersymmetric Poincare group on supersymmetric functions (see \cite{W},p.91, relation (14.22)).  \\
To close this section we formulate the invariance of super functions and super distributions of several variables by means of the generators $P,Q ,\bar Q $ of the translation group. This is needed in the next section. We restrict ourselves to a function or distribution
$F(z_1 ,z_2 )$ of two variables $z_1 =(x_1 ,\theta_1 ,\bar \theta_1 ), z_2 =(x_2 ,\theta_2 , \bar \theta_2 ) $. Let $P, Q_i ,\bar Q_i , i=1,2 $ be supersymmetric generators acting on the variables $z_1 $ and $z_2 $ respectively. We say that the function or distribution $F(z_1 ,z_2 )$ is supersymmetric translation invariant if
\begin{gather}
(P_1 +P_2)F(z_1 ,z_2 )=0  \\
(Q_1 +Q_2)F(z_1 ,z_2 )=0  \\
(\bar Q_1 +\bar Q_2)F(z_1 ,z_2 )=0  
\end{gather}
Instead of (11.8),(11.9) we may adopt
\begin{gather}
(D_1 +D_2)F(z_1 ,z_2 )=0  \\
(\bar D_1 +\bar D_2)F(z_1 ,z_2 )=0  
\end{gather}
by adopting the right instead of left multiplication \cite{WB} p.26 (in this case the covariant derivatives $D, \bar D $ and the charge operators $Q, \bar Q $ are interchanged).
The formal motivation of these definitions is obvious.

\section{Two point functions of quantized supersymmetric quantum field theory}

In this section we look for applications of the material exposed in the preceding sections to supersymmetric quantum field theory. First let us remark that we have explicitly constructed al least one example of a Hilbert space realized on supersymmetric functions which accommodates the symmetry group generators as sound operators. It may serve as an example of the Hilbert space which must be postulated in rigorous (relativistic) quantum field theory and as framework for studying such resistant subjects as canonical supersymmetric quantization. At the first glance canonical quantization in supersymmetry is hampered by the presence of so called auxiliary fields which seem to be non-quantizable because they are non-propagating fields. Based on the Krein-Hilbert structure it was possible to show that this is not the case at least at the level of canonical commutation relations \cite{C2}. \\
Here we present another application reaching free but also interacting fields which could be of interest. It is related to the celebrated K\"allen-Lehmann representation. The subject was already touched in \cite{C2} but some terms in the representation were missed.\\
Suppose that general principles of quantum field theory defined in Hilbert space \cite{SW} survive in the supersymmetric setting up \cite{O,C1,C2}. Then the two point function $W(z_1 ,z_2 )$ of a scalar neutral (or even complex) quantum field must satisfy the following requirements: \\
i) it must be a superdistribution (i.e. it has singularities) \\
ii) it must be invariant under the super Poincare group \\
iii) it must be positive definite \\
iv) it must satisfy $\overline{W(z_1 ,z_2 )}=W(z_2 ,z_1 )$ \\
The question is to find general $W(z_1 ,z_2 )$ satisfying i)-iv). Let us discuss the first requirement. We use a cheap definition of superdistributions (in two variables) requiring distribution coefficients in the series expansion in the Grassmann variables. Definitions using duality of linear locally convex spaces of test functions with appropriate topology are possible but will be not considered here (a natural system of seminorm can be given using \cite{R}). The second requirement on $W$ is
\begin{gather}
(Q_1 +Q_2)W(z_1 ,z_2 )=0  \\
(\bar Q_1 +\bar Q_2)W(z_1 ,z_2 )=0  
\end{gather}
where for the moment we leave out (11.7). Using (4.5),(4.6) this is a system of differential equations in the supersymmetric context. The reader can solve it easily by going to the new variables $\theta=\frac{1}{2}(\theta_1 +\theta_2 ), \zeta =\theta_1 -\theta_2 $ together with their conjugates $\bar \theta =\frac{1}{2}(\bar \theta_1 +\bar \theta_2 ), \bar \zeta =\bar \theta_1 -\bar \theta_2 $ as well as to $ x=x_1 -x_2 $ by translation invariance. The result is \cite{C2}
\begin{equation}
W(x,\theta ,\bar \theta ,\zeta ,\bar \zeta )=\mbox{ exp }[-i(\zeta \sigma^l\bar \theta -\theta \sigma^l\bar \zeta
)\partial_l ]E(x,\zeta ,\bar \zeta )
\end{equation}
where by invariance
\begin{gather}\nonumber
E(x,\zeta ,\bar \zeta )=E_1(x)+\zeta^2 E_2(x)+\bar \zeta^2E_3(x)+ \\ 
+\zeta \sigma^l\bar \zeta \partial_l E_4(x)+\zeta^2 \bar \zeta^2 E_5(x)
\end{gather}
with Lorentz invariant distributions $E_i =E_i (x_1 -x_2 ), i=1,2,...,5 $. The same conclusion follows if we adopt (11.10),(11.11) instead of (11.8),(11.9) (and replace $D$ operators by $Q$ operators and vice versa) but with 
\begin{gather}\nonumber
\mbox{ exp }[-i(\zeta \sigma^l\bar \theta -\theta \sigma^l\bar \zeta
)\partial_l ]
\end{gather}
in (12.3) replaced by
\begin{gather}\nonumber
\mbox{ exp }[i(\zeta \sigma^l\bar \theta -\theta \sigma^l\bar \zeta
)\partial_l ]
\end{gather}
Explicit computations can be found in \cite{C2} (see also \cite{GS} for a similar reasoning but in a different context). As far as $E_i $ are concerned it is well known that Lorentz invariant distributions are Fourier transforms of invariant measures in momentum space (spectral measures) of slow increase concentrated in the light cone. \\  
In this way we obtain a total of five linear independent contributions to the two point function which are supersymmetric invariant. On the other hand, from the investigations of the preceding sections there are five linearly independent explicitly known invariant kernels
\begin{gather}\nonumber
P_i K_i (z_1 -z_2 ), \quad i=c,a,T,+,-
\end{gather}
with $K_i (z_1 -z_2 )$ Lorentz invariant distributions multiplicated by $\delta^2 (\theta_1 -\theta_2 )\bar \delta^2 (\bar \theta_1 -\bar \theta_2 )$. It follows that $ W(z_1 ,z_2 )=W(x,\theta ,\bar \theta ,\zeta ,\bar \zeta )$ can be considered as a superposition of kernels of type we already studied in this paper. We get for  $W(z_1 ,z_2 )$: 
\begin{gather}
W(x_1 ,\theta_1 ,\bar \theta_1 , x_2 ,\theta_2 ,\bar \theta_2 )= \sum_i \lambda_i P_i K_i (z_1 -z_2 )
\end{gather}
where $\lambda_i $ are arbitrary complex parameters and we sum over $i=c,a,T,+,- $. Note that on the r.h.s. $P_i $ applied to $K_i (z_1 -z_2 )$ induces a $z_1 ,z_2 $ dependence not necessarily of the form $z_1 -z_2 $. By the fourth condition we must have as in Section 8 $ K_c =K_a =K_{c/a},K_+ =K_- =K_{\pm }$ and $\lambda_c =\lambda_a =\lambda_{c/a},\lambda_+ =\lambda_- =\lambda_{\pm } $. \\
For the convenience of the reader we give the explicit formulas which establish the connection between the contributions in (12.4) proportional to $E_i $ and $P_i $ applied to $\Delta =\delta^2 (\theta_1 -\theta_2 )\delta^2 (\bar \theta_1 -\bar \theta_2 )$ (do not confuse with $\Delta =\Delta (x)$ which appears in Sections 9 and 10). \\ 
Let
\begin{gather}\nonumber
S_1 =\zeta^2 e^{L} \\ \nonumber
S_2 =\bar \zeta^2 e^{L} \\ \nonumber
S_3 =e^{L} \\ \nonumber
S_4 = \zeta\sigma^l \partial_l \bar \zeta e^{L} \\ \nonumber
S_5 =\zeta^2 \bar \zeta^2 e^{L}=\zeta^2 \bar \zeta^2
\end{gather}
with
\[L=i(\theta \sigma^l \bar \zeta -\zeta \sigma^l \bar \theta )\partial_l =i(\theta_2 \sigma^l \bar \theta_1 -\theta_1 \sigma^l \bar \theta_2 )\partial_l \]
Then we have by computation
\begin{gather}\nonumber
S_1 =-\sqrt \square P_+ \Delta \\ \nonumber
S_2 =-\sqrt \square P_- \Delta \\ \nonumber
S_3 =\frac{\square}{4}(P_c +P_a -P_T )\Delta \\ \nonumber
S_4 =-\frac{i\square }{2}(P_c -P_a )\Delta \\ \nonumber
S_5 =\Delta=(P_c +P_a +P_T )\Delta
\end{gather}
Indeed the representations of $S_1 ,S_2 $ follow from the first two relations (9.41) p.73 in \cite{WB}. The expression for $S_3 $ follows from p.74 in \cite{WB}. The relation regarding $S_5 $ is trivial. It remains to prove that
\[ S_4 =-\frac{i\square }{2}(P_c -P_a )\Delta =\frac{i}{32}(D^2 \bar D^2 -\bar D^2 D^2 )\Delta    \]
This can be done by writing $\bar D^2 D^2 \Delta, D^2 \bar D^2 \Delta $ from (9.41) p.73 \cite{WB} in terms of $\theta =\frac{1}{2}(\theta_1 +\theta_2 ), \zeta =\theta_1 -\theta_2 $. It is a matter of long but elementary computations. \\
It remains to pass to the third condition concerning positivity. The positivity question can be partially answered as in Section 9. We get a positive definite kernel if we require $\lambda_{\pm}=0 $ , positive $\lambda_{c/a} $ and  $\lambda_T $ as well as the measures $d\rho $ (see Section 7) concentrated in the forward light cone. This was already noted in \cite{C2}. Certainly no condition relating $K_{c/a}$  to $K_T $ is necessary. \\
But there is a new interesting point which appears. Indeed it turns out that $P_+ ,P_- $ do not necessarily destroy  positivity, making $\lambda_{\pm }\neq 0 $ possible. We will show in this section that this is the case by dominating the contribution from $P_+ +P_- $ by the contribution from $P_c +P_a $.
The simplest idea would be to compute explicitly
\begin{gather}
\int d^8 z_1 d^8 z_2 \bar X(z_1 )[\lambda_{c/a}(P_c +P_a )K_{c/a} +\lambda_{\pm}(P_+ +P_- )K_{\pm }] 
 X(z_2 )
\end{gather}
by the methods used in the second proof of indefiniteness in Section 9 and to inquire positivity. But we prefer to return to the first proof of indefinite metric in Section 8 and split the problem into independent sectors: chiral and antichiral on one side and transversal on the other side, $X=X_c +X_a +X_T $. In order to start we compute besides $I_c $ in (8.10) also $I_+ $ for $X_c $ arbitrary chiral given in (8.8), (4.16) as well as $I_a ,I_- $ for $X_a $ antichiral. Recall that due to the fact that we compute integrands we can transfer freely space-time derivatives between factors. We have as in Section 8
\begin{gather} \nonumber
I_c (x_1 ,x_2 )=\int d^2 \theta d^2\bar \theta \bar X_c (x_1 ,\theta ,\bar \theta  )X_c (x_2 ,\theta ,\bar \theta )= \\ 
=\bar f_c (x_1 )\square f_c (x_2 )-\frac{i}{2}\bar \varphi_c (x_1 )\bar \sigma^l \partial_l \varphi_c (x_2 )+\bar m_c (x_1 )m_c (x_2 )  \\ \nonumber
I_a (x_1 ,x_2 )=\int d^2 \theta d^2\bar \theta \bar X_a (x_1 ,\theta ,\bar \theta )X_a (x_2 ,\theta ,\bar \theta )= \\ 
=\bar f_a (x_1 )\square f_a (x_2 )-\frac{i}{2}\chi_a (x_1 )\sigma^l \partial_l \bar \chi_a (x_2 )+\bar n_a (x_1 )n_a (x_2 )  
\end{gather} 
as well as 
\begin{gather} \nonumber
I_+ (x_1 ,x_2 )=\frac{1}{4\sqrt {\square }}\int d^2 \theta d^2 \bar \theta \bar X_a (x_1 ,\theta ,\bar \theta )D^2 X_c (x_2 ,\theta ,\bar \theta ) = \\ 
=-\bar f_a (x_1 )\sqrt {\square } m_c (x_2 )-\bar n_a (x_1 )\sqrt {\square }f_c (x_2 )+\frac{1}{2}\chi_a (x_1 )\sqrt {\square }\varphi_c (x_2 ) \\ \nonumber
I_- (x_1 ,x_2 )=\frac{1}{4\sqrt {\square }}\int d^2 \theta d^2 \bar \theta \bar X_c (x_1 ,\theta ,\bar \theta )\bar D^2 X_a (x_2 ,\theta ,\bar \theta )= \\ 
=-\bar f_c (x_1 )\sqrt {\square } n_a (x_2 )-\bar m_c (x_1 )\sqrt {\square }f_a (x_2 )+\frac{1}{2}\bar \varphi_c (x_1 )\sqrt {\square }\bar \chi_a (x_2 )
\end{gather} 
The idea is to dominate $ \lambda_{\pm}(I_+ +I_- ) $ by $ \lambda_{c/a }(I_c +I_a ) $. 
We start by studying the case  
\begin{gather}
K_{\pm }=K_{c/a}
\end{gather}
This is an extra condition which in physics could be motivated by requiring same mass spectrum for all components of the supersymmetric multiplet. But we will eliminate it at the end of the paper.
If (12.11) holds than the reader can convince himself using (12.7)-(12.10) that the positivity requires (beside $ \lambda_{c/a}>0 $)
\begin{gather}
-\lambda_{c/a}<\lambda_{\pm} < \lambda_{c/a} 
\end{gather}  
Finally the transversal contribution doesn't interact with other contributions and from Section 8 as well as from Section 9 it follows that positivity imposes $\lambda_T >0 $. \\
We collect the results into the K\"allen-Lehmann representation for the free and interacting supersymmetric scalar field: \\
The general two point function of the scalar neutral (or even complex) supersymmetric field has the representation 
\begin{gather} \nonumber
W(z_1 ,z_2 )=(\lambda_{c/a}(P_c +P_a )+\lambda_{\pm}(P_+ +P_- ))K_{c/a} (z_1 -z_2 )- \\ 
-\lambda_T P_T K_T (z_1 -z_2 )
\end{gather}
where the $\lambda $-parameters are restricted to
\begin{gather}
\lambda_{c/a},\lambda_T >0, \quad
-\lambda_{c/a}<\lambda_{\pm} <\lambda_{c/a} 
\end{gather}
and $K_{c/a} ,K_T $ are of the form (5.31),(5.32). \\
If we do not assume
\begin{gather}\nonumber
K_{\pm }(z)=K_{c/a}(z)
\end{gather}
then the result (12.13) changes only minimally. In this case positivity restricts the general two point K\"allen-Lehmann representation to
\begin{gather} \nonumber
W(z_1 ,z_2 )=(\lambda_{c/a}(P_c +P_a ))K_{c/a} (z_1 -z_2 )+(\lambda_{\pm}(P_+ +P_- ))K_{\pm } (z_1 -z_2 )- \\ 
-\lambda_T P_T K_T (z_1 -z_2 )
\end{gather}
where (12.14) has to be replaced by
\begin{gather}
\lambda_{c/a},\lambda_T >0 \\
-\lambda_{c/a}\rho_{c/a }(p)<\lambda_{\pm}\rho_{\pm }(p)<\lambda_{c/a}\rho_{c/a }(p) 
\end{gather}
In (12.17) $\rho_{c/a}(p)$ and $\rho_{\pm}(p)$ are the densities of the measures which appear in the Fourier transform (5.32) for $K_{c/a}$ and for $K_{\pm}$ respectively. In fact (because in this section the Lorentz invariance is implicit) these measures depend only of $p^2 $. The inequalities (12.17) should hold for all values of the momentum $p$. The condition (12.17) follows from the positivity by restricting the coefficients of $ X_c ,X_a $ to an arbitrary small neighborhood of a given momentum $p$. The representation (12.15) could be simplified by absorbing the positive $\lambda $-coefficients in $K $ (and the measures $d\rho $).\\
Finally note that using the methods of this paper it is possible to write down a two by two matrix K\"allen-Lehmann representation for models \cite{WB} of Wess-Zumino type too. The problem is even simpler because the transversal sector in not involved. The (matrix) domination of $P_+ ,P_- $ by the $ P_c ,P_a $ contributions is similar. The supersymetric free two point functions \cite{WB,C3} are particular cases of (12.13). \\

At the end of this section and at the interface between mathematical and physical considerations, let us add some comments and mention at the same time some perspectives of the present work. The positive bilinear form (produced by the two point function) is strictly positive definite if it is induced by the kernel studied in Section 9:
\begin{gather}
JK_0 =(P_c +P_a -P_T )K_0
\end{gather}
This is particularly interesting if we try to connect to the classical Bochner-Schwartz theorem of distribution theory \cite{GV}. In this classical context multiplicatively positive definite bilinear forms are characterized by positive tempered measures. Certainly the measure theory collapses in the supersymmetric framework. But the situation is not as bad as it appears to be. First of all let us remark that for the supersymmetric results of this section we used Poincare supersymmetry which implies Lorentz invariance. But it can be shown \cite{C4} that full Poincare supersymmetry is not needed; invariance under the supersymmetric translation group is sufficient. This would imply a Bochner-Schwartz theorem for positive definite supersymmetric translation invariant bilinear forms. The measure-theoretic framework has to be modified; more precisely it has to be enriched by the supersymmetric projections as this was worked out in this paper. Returning to the classical case, the Bochner-Schwartz theorem is connected to the famous Bochner theorem which can be used in order to study unitary representations of the translation group (Stone). Now the idea is to use the supersymmetric Bochner-Schwartz theorem in order to study the supersymmetric counterpart of the Stone (or even SNAG) theorem (related to the supersymmetric translation group). The point is that measure-theoretic aspects do not collaps completely and presumably the "spectral projections" of the classical Stone theorem have to be enriched by exactly the supersymmetric projections $P_c ,P_a ,P_T $. Besides this the only new aspect should be the Krein structure of this paper.\\ 
Acknowledgements: \\
We thank K.H. Rehren, G.M. Graf and M. Schork for correspondence which helped improving the paper.


\begin{thebibliography}{99}


\bibitem{C1} F. Constantinescu, arXiv:0305143, J.Phys.A: Math.Gen. 38(2005),1385; 39(2006),9903
\bibitem{WB} J. Wess, J. Bagger, Supersymmetry and supergravity, 2nd edition, Princeton University Press, 1992 
\bibitem{DeW} B. DeWitt, Supermanifolds, Cambridge University Press, Cambridge, 1992
\bibitem{DM} P. Deligne, J.W. Morgan, Notes on supersymmetry, quantum fields and strings: a course for mathematicians, vol.1,2 Amer. Math. Soc., Providence RI, 1999, 41-97
\bibitem{V} V.S. Varadarajan, Supersymmetry for mathematicians: an introduction, Courant lecture notes 11, American Mathematical Society, Providence, Rhode Island, 2004
\bibitem{StW} F. Strocchi, A.S. Wightman, Journ.Math.Phys. 15(1974),2198
\bibitem{St} F. Strocchi, Selected topics on the general properties of quantum field theory, World Scientific, 1993
\bibitem {SW} R.F. Streater, A.S. Wightman, PCT, spin and statistics and all that, Benjamin, 1964
\bibitem{Sch} G. Scharf, Quantum gauge theory-a true ghost story, Wiley Interscience, 2001
\bibitem{G} D.R. Grigore, Romanian Journ.Phys. 44(1999),853
\bibitem{S} P.P. Srivastava, Supersymmetry, superfields and supergravity: an introduction, IOP Publishing,
Adam Hilger, Bristol, 1986
\bibitem{RY} W. R\"uhl, B.C. Yunn, Fortsch.Phys. 23(1975),431; 23(1975),451
\bibitem{I} E.A. Ivanov, Supersymmetry at BLTP: how is started and where we are, hep-th/0609176
\bibitem{H} M. Henneaux, C. Teitelboim, Quantization of gauge systems, Princeton University Press, Princeton, New Jersey, 1992
\bibitem{C2} F. Constantinescu, Intern.Journ.Modern Physics 21(2006),2937; Annalen Phys. 15(2006),861
\bibitem{O} K. Osterwalder, Supersymmetric quantum field theory, in V. Rivasseau (ed.), Results in field
theory, statistical mechanics and condensed matter physics, Lecture Notes in Phys. 446, Springer, New York, 1995,
117
\bibitem {W} P. West, Introduction to supersymmetry and supergravity, Extended second edition, World Scientific, 1990
\bibitem {R} A. Roger, Supermanifolds, World Scientific, 2007
\bibitem{GS} D.R. Grigore, G. Scharf, Annalen Phys. 12(2003),5
\bibitem{C3} F. Constantinescu, Lett.Math.Phys. 62(2002),111
\bibitem{GV} I.M. Gel'fand, N.Ya. Vilenkin, Generalized functions, vol 4, Academic Press, 1964
\bibitem{C4} F. Constantinescu, work in progress
\end{thebibliography}
\end{document}